\documentclass[aps,prl,showpacs,reprint,groupedaddress,amsmath,amssymb, floatfix]{revtex4-1}
\usepackage{graphicx}
\usepackage{epstopdf}
\usepackage{subfigure}
\usepackage{dcolumn}
\usepackage{bm}
\usepackage{extarrows}
\usepackage{hyperref}
\hypersetup{colorlinks=true,       
   linkcolor=blue,          
   citecolor=blue,        
   urlcolor=blue }	

\def\be{\begin{equation}}      
\def\ee{\end{equation}}
\def\beu{\begin{equation*}}   
\def\eeu{\end{equation*}}

\providecommand{\ket}[1]{\left|#1\right\rangle}

\providecommand{\mean}[1]{\left\langle#1\right\rangle}

\begin{document}
\title{Quantum Nonlinear Optics Near Optomechanical Instabilities}


\author{Xunnong Xu }
\author{Michael Gullans}
\author{Jacob M. Taylor}
\affiliation{Joint Quantum Institute, University of Maryland/National Institute of Standards and Technology, College Park, Maryland 20742, USA}



\date{\today}
\begin{abstract}
Optomechanical systems provide a unique platform for observing quantum behavior of macroscopic objects. However, efforts towards realizing nonlinear behavior at the single photon level have been inhibited by the small size of the radiation pressure interaction. Here we show that it is not necessary to reach the single-photon strong-coupling regime in order to realize significant optomechanical nonlinearities.  Instead,  nonlinearities at the few quanta level can be achieved, even with weak-coupling, in a two-mode optomechanical system driven near instability. In this limit, we establish a new figure of merit for realizing strong nonlinearity which scales with the single-photon optomechanical coupling and the sideband resolution of the mechanical mode with respect to the cavity linewidth.  We find that current devices based on optomechanical crystals, thought to be in the weak-coupling regime, can still achieve strong quantum nonlinearity; enabling deterministic interactions between single photons.
\end{abstract}


\maketitle

Recent years have seen dramatic progress in realizing deterministic interactions between single photons, which has profound implications for future optical technologies \cite{Milburn89, Turchette95, Caulfield10,Obrien09}.  The most striking success has been achieved with cavity quantum electrodynamics (cQED) \cite{Birnbaum2005, Press07,Bishop2009,Volz2012,Bose12,Chen13,Reiserer14,Tiecke14}, where  photons inherent the saturation of a single two-level atom due to strong interactions between the atom and the cavity field.  Alternative approaches have been explored based on slow-light-enhanced Kerr nonlinearites \cite{Imamoglu1997,Lukin00,Lin12}, single dye-molecules \cite{Hwang09}, strong photon interactions mediated by Rydberg atoms  \cite{Peyronel2012,Dudin12,Gorn14,Tiarks14},  enhanced nonlinearities in plasmonic systems \cite{Chang07,Gullans2013} and atoms coupled to wave guides\cite{Bajcsy09,Venk11, Kolchin2011,OShea13}. 

Optomechanical systems, where light and mechanical motion are coupled by radiation pressure \cite{Kippenberg2008, Aspelmeyer2013, Thompson08, Chan2011, Verhagen2012, Groblacher2009, Teufel2011}, are a promising approach to realizing strong photon interactions.  Unfortunately no experiment has yet managed to reach the single-photon strong coupling regime.  Recently it was noted that, in the weak coupling regime, there are still signatures of optomechanical nonlinearity \cite{Lemonde2013,Borkje2013,Kronwald2013}; however, strong coupling is required to achieve significant nonlinear quantum effects and deterministic photon interactions with optomechanics \cite{Rabl2011, Nunnenkamp11, Qian12, Komar13}.

In this Letter, we show it is not necessary to reach the quantum strong coupling regime in order to obtain large single-photon nonlinearities.  Instead, in two-mode optomechanical systems with strong side-band resolution, the nonlinearity can be enhanced to the single-photon level by driving the system near an instability.  In particular,
as the strength of the driving field increases, the frequency of one of the optomechanical normal modes approaches zero and the associated harmonic oscillator length becomes large \cite{Lu2013}. The increased quantum fluctuations associated with this mode result in an enhanced nonlinear interaction.  We show that when the mechanical mode is sideband resolved with respect to the cavity, the enhancement  in the nonlinear coupling can exceed the dissipation by an amount scaling with the sideband resolution $\omega_m/\kappa$, where $\omega_m$ is the mechanical frequency and $\kappa$ is the cavity linewidth.  We demonstrate that this results in enhanced photon-photon interactions by calculating the equal time, two-photon correlation function $g^{(2)}(0)$ for weakly incident probe light.  The presence of anti-bunching $g^{(2)}(0)<1$ in the cavity output field indicates the onset of photon blockade and, in this case, significant two-photon nonlinearity. We inferred  a new parameter $P = g_0^2\omega_m/\kappa^3$ ($g_0$ is single-photon optomechanical coupling), whose largeness is the relevant quantity for determining the strength of the nonlinearity. We find that in current devices based on optomechanical crystals, our approach could increase the observable antibunching by more than an order of magnitude.

\begin{figure}[t]
\begin{center}
\includegraphics[width=0.49\textwidth]{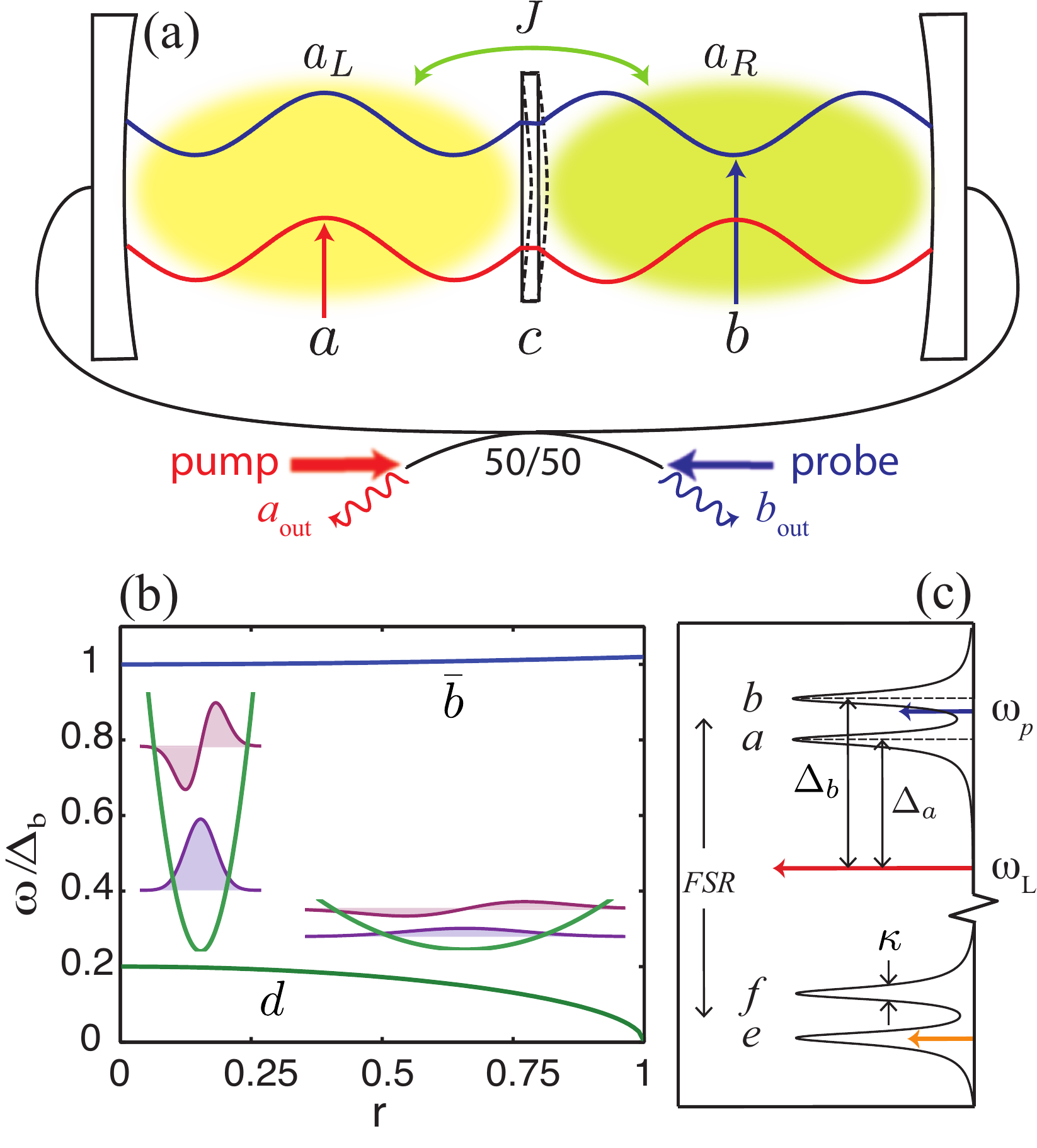}
\caption{(color online). (a) Schematic of the coupled two-mode system. Displacements of the middle mirror (via mechanical oscillations) couple the symmetric mode $a$ (red) and antisymmetric mode $b$ (blue) as the left-right symmetry is broken. (b) Normal modes of the coupled harmonic oscillator bilinear hamiltonian for $\Delta_b=5 \omega_m$, with blue (green) line representing the higher (lower) energy branch $\bar{b}$ ($d$). As pump power increases, the energy of the lower branch decreases, the effective potential becomes flat and the associated harmonic oscillator length becomes larger. 
(c) Energy scales for the pump, probe and cooling modes.} 
\label{fig:fig1}
\end{center}
\end{figure}

The system we consider is shown in Fig.~\hyperref[fig:fig1]{1(a)}.  It consists of a high finesse optical cavity that has two spatially separated, degenerate optical modes ($a_L, a_R$) at frequency $\omega_c$ coupled at a rate $J$ through a mirror with near perfect reflection \cite{ Xu2013}.  Both optical modes are also coupled to a common mechanical mode ($c$)  through radiation pressure with single-photon optomechanical coupling rate $g_0$. In the symmetric-antisymmetric mode basis $a=(a_L+a_R)/\sqrt{2}$, $b=(a_L-a_R)/\sqrt{2}$ the Hamiltonian is ($\hbar =1$):
\begin{align} \nonumber
H &= (\omega_c - J) a^{\dagger}a + (\omega_c +J) b^{\dagger}b + \omega_m c^\dagger c \\
& -  g_0 (a^{\dagger}b + b^{\dagger}a)(c+c^\dagger)\ . 
\end{align}
In addition, there is also a dissipative interaction of the cavity and mechanical modes with their environment, with a conservative term $V =  \sqrt{\kappa} (a_{in}(t) a^\dagger+h.c.)$ and damping $\kappa$ (described below).
The two cavities are assumed to have identical damping rates, while $a_{in}$ is the input fields for the symmetric mode.  

In the presence of a strong  drive $a_{in}(t)=a_{in}+\sqrt{\kappa}\, \alpha_p e^{-i \omega t}$ there is an an effective linear coupling between the antisymmetric mode and the  mechanical mode, and also a residual nonlinear coupling between the mechanical mode and both optical modes.  The Hamiltonian in the rotating frame for the pump displaced oscillator states ($a\to a+\alpha$) becomes \cite{Aspelmeyer2013}
\begin{align}
H  &=  \Delta_a a^{\dagger}a + \Delta_b b^{\dagger}b + \omega_m c^\dagger c -  G_0 (b+b^{\dagger}) (c+c^\dagger)   \nonumber \\
&-  g_0(a^{\dagger}b + b^{\dagger}a) (c+c^{\dagger}) 
\end{align}
where $\Delta_{a(b)}$ is the detuning of mode $a$ ($b$) with respect to the pumping laser and $G_0 \equiv g_0 \alpha = g_0 \alpha_p \kappa/(\Delta_a-i\kappa/2)$ is the pump-enhanced linear coupling. By choosing an appropriate phase of the pump, we can make $G_0$ real. In what follows, we make the further assumptions that $\Delta_b\gg \omega_M$, such that the parameter $\eta \equiv \omega_m/\Delta_b$ is much smaller than 1.  In this regime the hybridized polariton modes Eq.~(\ref{eqn:polariton1})-(\ref{eqn:polariton2}) retain mostly their original photonic or mechanical character, reducing the deleterious effect of optical loss on the `mechanical' mode.  We give the full expressions in the supplementary material \cite{supp}.

The first four terms in $H$ are bilinear in the oscillator modes and can be  diagonalized to give the normal modes
\begin{align} \label{diagH}
H_0  =\Delta_a a^{\dagger}a +  (\Delta_b+\delta)  \bar{b}^{\dagger} \bar{b} +    \omega_m \zeta\, d^{\dagger}d, 
\end{align}
with the normal mode frequencies given in terms of the parameters $\delta \approx r^2 \omega_m\eta/2$ and $\zeta = \sqrt{1-r^2}$ to first order in $\eta$.  We defined the rescaled driving amplitude $r\equiv 2 G_0/\sqrt{\omega_m \Delta_b}$.  As $r\to1$ the frequency of the lower branch goes to zero and the mode effectively becomes a free particle, leading to enhanced quantum fluctuations in this mode, as shown in Fig.~\hyperref[fig:fig1]{1(b)}.  For $r>1$, the normal mode frequency becomes imaginary signifying the onset of the instability.  
For  $0\le r<1$ and $\eta \ll1$, the normal mode operators are, surprisingly 
\begin{align}
\bar{b} &\approx  b - \frac{r}{2}\sqrt{\eta} (c+c^\dagger), \label{eqn:polariton1} \\ 
d &\approx \frac{1}{2 \sqrt{\zeta}}(c-c^\dagger)+ \frac{\sqrt{\zeta}}{2} (c+c^\dagger)   + \frac{r}{2}\sqrt{\eta}(b-b^\dagger). 
 \label{eqn:polariton2}
\end{align} 
In this regime, $\bar{b}$ is mostly optical while $d$ is mostly mechanical, to $O(r\sqrt{\eta})$. Note that $d$ represents a squeezed state in the quadrature variables for small $\zeta$ as explained in \cite{supp}. 
Including the nonlinearity, we can reexpress the normal-ordered Hamiltonian to first order in $\eta$
\begin{align} \label{eqn:nlenhanced}
H&=H_0 - \frac{g_0}{\sqrt{\zeta}} (a^\dagger \bar{b} + a \bar{b}^\dagger)(d+d^\dagger) \\ \nonumber
&-\frac{g_0}{\sqrt{\zeta}} \sqrt{\frac{\eta}{4\zeta}} (a+a^\dagger)(d^2+d^{\dagger 2} +2 d^\dagger d). 
\end{align}
Near the instability, $\zeta \ll 1$, the effective optomechanical coupling $g_0/\sqrt{\zeta}$ is strongly enhanced.  We remark that this approach is distinct from simply choosing a low frequency mechanical oscillator to begin with.  In particular, the mass and frequency of a mechanical oscillator (of the same shape and material) are usually related to each other by $\omega_m \propto \sqrt{1/m}$, so that the stiffness $m \omega_m^2$ is roughly the same for different oscillators. This suggests that higher frequency oscillators have larger intrinsic position fluctuation than low frequency ones, since $x_{\mathrm{zpf}} = \sqrt{\hbar/m\omega_m}$, which further implies larger optomechanical coupling $g_0$. This back-action induced softening of harmonic oscillator has the benefits of combining small mass and low frequency, so the effective coupling can be enhanced substantially.   

To utilize this enhanced nonlinear coupling at the single quanta level we need to consider the effects of both dissipation and terms in Eq.~(\ref{eqn:nlenhanced}) which tend to destabilize the system towards large mode occupation. 
In the normal mode basis, $H$ contains five distinct nonlinear interactions:
\be \label{eqn:nlterms}
\begin{array}{c c c}
\bar{b}^\dagger a d + h.c., & a^\dagger \bar{b} d + h.c., & a^\dagger d d + h.c., \\
 a d d + h.c., & (a+a^\dagger)d^\dagger d. &
\end{array}
\ee
When the frequency of the $d$ mode is small, these nonlinear terms will destabilize the system towards large mode occupation, which, together, with the cavity induced decay will contaminate any few photon effects.  To keep the system far in the stable regime, we require $g_0/\sqrt{\zeta},g_0 \sqrt{\eta/\zeta^2} \ll \omega_m \zeta$, which further constrains $\Delta_b$ and $\zeta$.  In addition, the mechanical mode must be close to the ground state, below we show how this can be achieved with optomechanical cooling for the normal mode when the heating rate $\gamma_\uparrow$ (defined below) is much less than $\kappa$. In particular, to have a large effective single photon optomechanical nonlinearity we have to satisfy
\be \label{eqn:ineq}
\gamma_{\uparrow} \ll \kappa \ll g_0/\sqrt{\zeta}\ll \omega_m \zeta. 
\ee
It is easy to show that this can be satisfied for large $\Delta_b$ and small $\gamma_\uparrow$ when 
\be \label{figure of merit}
P \equiv \frac{g_0^2}{\kappa^2} \frac{\omega_m}{\kappa} \gg 1.  
\ee
Thus the condition for strong optical nonlinearities is relaxed from $g_0 \gg \kappa$ to $P\gg1$.  Below we show that this enhanced nonlinearity can be used to achieve photon blockade.

We can describe dissipation with the master equation for the density matrix $\rho$ of the three-mode system
\begin{align} \nonumber
\dot{\rho} &= -i[H,\rho] -\kappa \big( \mathcal{D}[a] +\mathcal{D}[b]\big) \rho \\
&-\gamma_m\big((\bar{n}_{\mathrm{th}}+1) \mathcal{D}[c]+ \bar{n} \mathcal{D}[c^\dagger])\big)\rho, 
\label{mastereq}
\end{align}
where $\mathcal{D}[A]\rho = 1/2 \, \{A^\dagger A,\rho \} - A \rho A^\dagger$  for any operator $A$, $\gamma_m$ is the mechanical heating rate, and $\bar{n}_{\mathrm{th}}$ is the thermal occupation of the mechanical mode in the absence of the coupling to the cavity.
In the normal mode basis, the jump operator for the cavity and mechanical modes become $b\to \bar{b} + \sqrt{\eta/\zeta}\, \big(d+d^\dagger \big)/2$ and $c \to (d+d^\dagger)/2\sqrt{\zeta} + \sqrt{\zeta}(d-d^\dagger)/2$, respectively, implying that dissipation of both cavity mode $b$ and mechanical mode $c$ results in added noise on the $d$ mode.  Near the instability $\zeta \ll 1$, the downward transitions (emission)  and upward transitions (absorption) in the $d$ mode occur at the respective rates
\begin{align}
\gamma_{\downarrow} &=  \frac{\eta}{4 \zeta} \kappa+\frac{\gamma_m}{4 \zeta} (2 \bar{n}_{\mathrm{th}}+1+2\zeta), \\
\gamma_{\uparrow} &=  \frac{\eta}{4 \zeta} \kappa+\frac{\gamma_m}{4 \zeta} (2 \bar{n}_{\mathrm{th}}+1-2\zeta). 
\end{align}
Since $\gamma_\downarrow-\gamma_\uparrow = \gamma_m \ll \gamma_\uparrow$, the absorption terms will tend to excite the $d$ mode to high occupation numbers roughly given by $\bar{n}_d \sim \gamma_\uparrow/(\gamma_\downarrow-\gamma_\uparrow)$ \cite{Aspelmeyer2013}.

A natural way to overcome this difficulty is to add optomechanical cooling to the $d$ mode. As shown in Fig.~\hyperref[fig:fig1]{1(c)}, we consider using another pair of cavity modes $e, f$ separated by the cavity free spectrum range (FSR) to induce sideband cooling of the $d$ mode. Driving mode $e$ enhances the coupling between mode $f$  and the mechanical mode $c$  by an amount $\alpha_e$, the steady state amplitude of $e$. Moving to the optomechanical normal mode basis, we get the additional terms in the hamiltonian: 
\be
\Delta_e e^{\dagger} e + \Delta_f f^{\dag}f - \frac{g_0}{\sqrt{\zeta}} \alpha_e (f+f^{\dag})(d+d^{\dag}).
\ee
We see that the coupling is further enhanced by $1/\sqrt{\zeta}$ because of the increase of harmonic oscillator length. Similar to the usual single-mode optomechanical cooling, when $\Delta_f = \omega_m\zeta$, the $d$ mode is cooled by the $f$ mode \cite{Aspelmeyer2013} and the system reaches steady state quickly.

\begin{figure}[t]
\begin{center}
\includegraphics[width=0.49\textwidth]{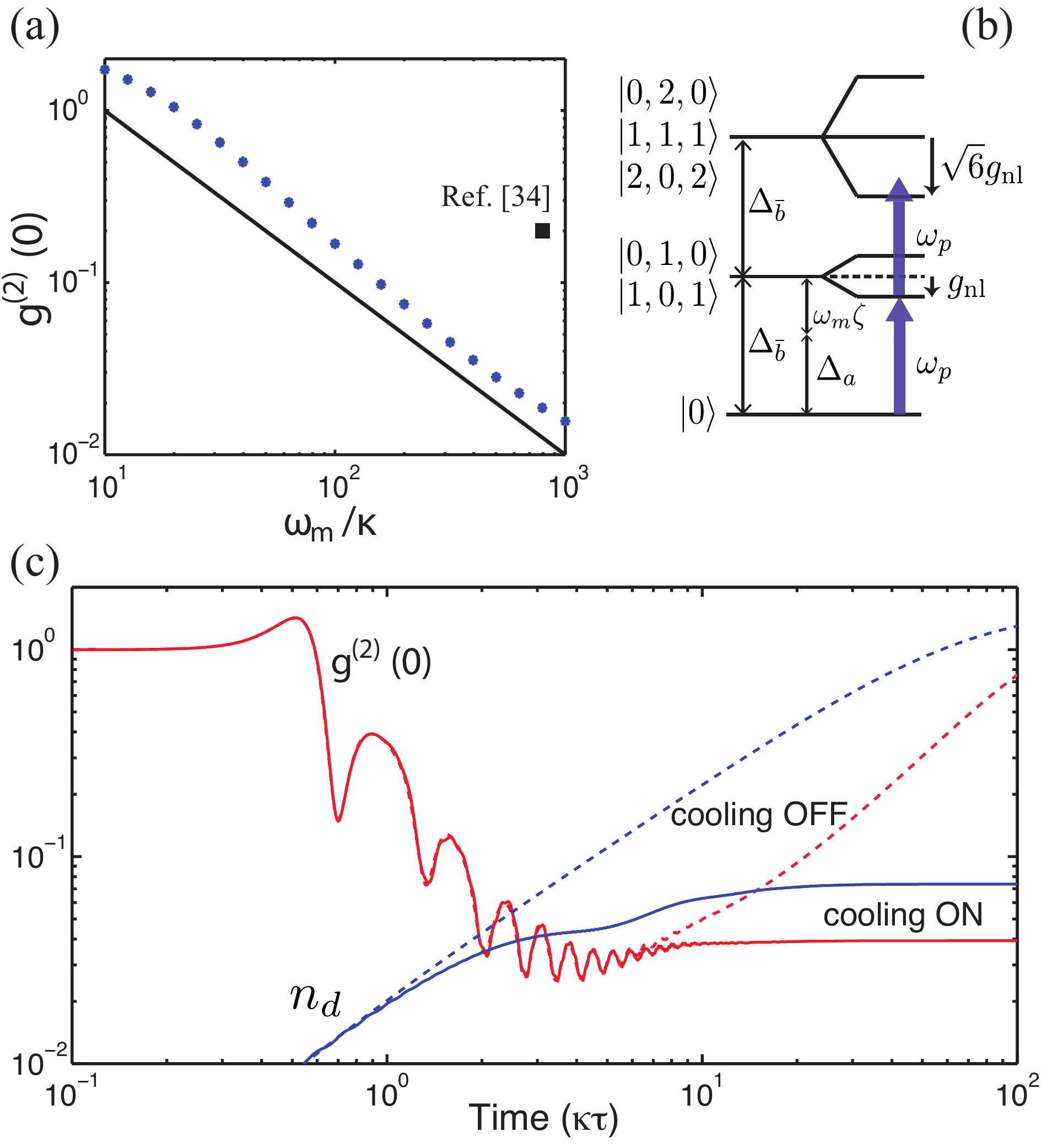}
\caption{(color online). (a) Dotted line shows $g^2(0)$ of the $\bar{b}$ mode as a function of $P$ with $g_0=\kappa$ (so $P=\omega_m/\kappa$), $\alpha_e = 0.1$ and a small coherent probe field in $\bar{b}$ with strength $\beta_{\bar{b}} = 0.02 \kappa$.  We restrict the mode occupations to be less than 4. When $P  > 40$, $g^2(0)$ roughly scales as $1/P$ (black line). The black square represents the value of $g^2(0)$ obtained in Ref.\ \cite{Lemonde2013} when $\omega_m/\kappa \to \infty$. For comparison, the $g^2(0)$ in Ref.\ \cite{Rabl2011}, when $g_0 = \kappa$, increases linearly with $P^2$.    (b)The level diagram of the system when the interaction $\bar{b}^{\dagger}ad$ becomes resonant. The probe field drives the lower energy state of the one-photon manifold on resonance. (c) Evolution of the equal time, two-photon correlation function $g^{(2)}(0)$ (red lines) and the population in the $d$ mode $n_d = \mean{d^\dagger d}$ (blue lines), after the probe field is turned on  ($P=500$, $g_0=\kappa$, $\gamma_{\uparrow\downarrow} = \kappa/\sqrt{P}$ and $\beta=0.02\kappa$). In addition, dotted lines show the result for $\alpha_e = 0$ indicating cooling on  $d$ is OFF, while solid lines are for $\alpha_e = 0.1$ indicating cooling is ON. }
\label{fig:fig2}
\end{center}
\end{figure}

The nonlinear terms will have the strongest effect when one of the interactions in Eq.~(\ref{eqn:nlterms}) is tuned into resonance: $\Delta_{\bar{b}} = \Delta_a + \omega_m \zeta$ for the $\bar{b}^\dagger a d$ term, $\Delta_a= \Delta_{\bar{b}} + \omega_m \zeta$ for the $a^\dagger \bar{b} d$ term, and $\Delta_a = 2 \omega_m \zeta$ for the $a^\dagger dd$ term, where $\Delta_{\bar{b}} = \Delta_b + \delta$ is the energy of the normal mode $\bar{b}$ and $\Delta_{\bar{b}}/\Delta_b =1$ up to first order in $\eta$.  The lower two interactions are always off-resonant and will tend to destabilize the system towards large mode occupation, but they will be suppressed when Eq.~(\ref{eqn:ineq}) is satisfied. Here we focus on the resonant interactions $\bar{b}^{\dagger}ad$ and $a^{\dagger}\bar{b}d$ because the $a^\dagger dd$ interaction coefficient is weaker.   
In the occupation number basis $\ket{n_a, n_{\bar{b}}, n_d}$,  the 2-fold degeneracy of the first excited state is broken by $g_{\mathrm{nl}}$ and the 3-fold degeneracy of the second excited state is broken by $\sqrt{6} g_{\mathrm{nl}}$ due to the 3-body interaction $\bar{b}^{\dagger}ad + \mathrm{h. c.}$:
\begin{align}  
&  \ket{0, 1, 0}\xlongleftrightarrow{ g_{\mathrm{nl}}} \ket{1,0,1},  ~\Delta\omega:   \pm g_{\mathrm{nl}} \nonumber \\
&  \ket{0, 2, 0}\xlongleftrightarrow{\sqrt{2} g_{\mathrm{nl}}} \ket{1,1,1}   \xlongleftrightarrow{2 g_{\mathrm{nl}}} \ket{2,0,2}, ~ \Delta\omega: 0,  \pm \sqrt{6} g_{\mathrm{nl}}  \nonumber 
\end{align} 
with $g_{\mathrm{nl}} = g_0/\sqrt{\zeta}$.  Since $\bar{b}$ has a strong overlap with the antisymmetric cavity mode, we can optically probe it as illustrated in Fig.~\ref{fig:fig1}.  Similar to the Jaynes-Cummings nonlinearity in cQED system \cite{Birnbaum2005}, when probing the $\bar{b}$ mode at frequency $\omega_p = \Delta_{\bar{b}} - g_{\mathrm{nl}}$ with strength $\beta_{\bar{b}}$, we can observe a photon-blockade effect because of the anharmonicity of the ladders, which is shown in Fig.~\hyperref[fig:fig2]{2(b)}.   The signature of the photon blockade will be in the antibunching of the output light, i.e., when $g^{(2)}(0) <1$, where $g^{(2)}(0)$ is the equal time, two-photon correlation function defined by 
\be
 g^{(2)}(t)=\frac{\mean{ \bar{b}^\dagger(\tau) \bar{b}^\dagger(\tau+t) \bar{b}(\tau+t) \bar{b}(\tau)}}{\mean{ \bar{b}^\dagger(\tau) \bar{b}(\tau)}^2}
 \label{eq:g2def}
\ee
for a given evolution time $\tau$.  Fig.~\hyperref[fig:fig2]{2(a)} shows that, for optimal parameters described below, the minimum value of $g^{(2)}(0)\sim 1/P$, thus the system exhibits a strong single photon nonlinearity even when $g_0 \lesssim \kappa$. 
We note that, it is the anti-symmetric mode $b = \bar{b} + \sqrt{\eta/\zeta}(d+d^{\dag})/2$ that actually comes out of the cavity, but, the contribution from the $d$ mode is suppressed for small $\eta/\zeta$.  In addition, the large frequency splitting between the $\bar{b}$ and $d$ mode enables the two contributions to be measured separately.  


Fig.~\hyperref[fig:fig2]{2(c)} shows the typical evolution of $g^{(2)}(0)$ with $\tau$ obtained from numerical simulation of the master equation.  Without cooling, after initial transient dynamics, the system reaches a quasi-steady state with strong antibunching.  Eventually, the system is pumped to states with a finite population in $d$ as shown by the dashed blue line in Fig.~\hyperref[fig:fig2]{2(c)}.  These states, $\ket{0,0,n}$, are dark states of the system for $n>0$, because, due to the nonlinearity, they are no longer resonantly excited by the $b$-probe.  As a result, the antibunching is reduced at long times. However, in the presence of cooling these dark states are depopulated and the system reaches a steady state with strong antibunching. 

\begin{figure}[t]
\begin{center}
\includegraphics[width=0.48\textwidth]{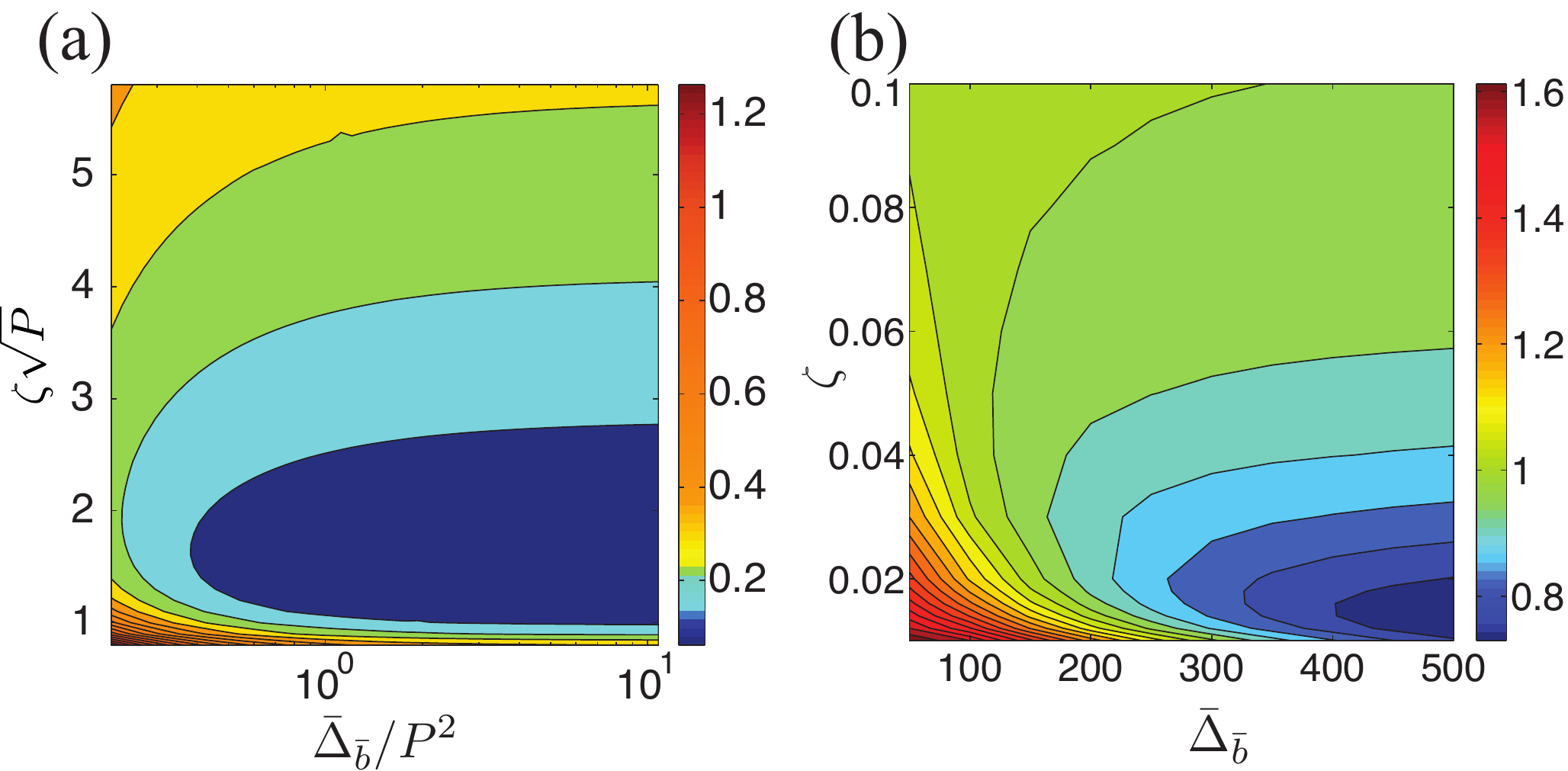}
\caption{(color online). Contour plots of the minimum $g^{(2)}(0)$ in steady state versus the experimental control parameters $\bar{\Delta}_{\bar{b}}=\Delta_{\bar{b}}/\kappa$ and $\zeta$. (a) $P=100, ~\omega_m/\kappa = P$;  (b) $g_0/\kappa = 0.1$, $\omega_m/\kappa = 500$, $P=5$ and $\alpha_e = 2\sqrt{\zeta}$.}
\label{fig:fig3}
\end{center}
\end{figure}

To achieve single photon blockade using the scheme illustrated in Fig.~\hyperref[fig:fig2]{2(b)}, we also need to satisfy the inequalities given in Eq.~(\ref{eqn:ineq}), which requires optimization of the system parameters. 
The original Hamiltonian has six independent parameter: $(\Delta_a, \Delta_b, \omega_m, G_0, g_0, \kappa)$, but rescaling by $\kappa$ and taking the resonance condition $\Delta_{\bar{b}}=\Delta_a +\omega_m \zeta$, we are left with four independent parameters: $(P, \omega_m/\kappa, \Delta_b/\kappa, \zeta)$. $P$ and $\omega_m/\kappa$ are device-dependent parameters we want to tune, while $\Delta_b/\kappa$ and $\zeta$ can be controlled by tuning the frequency and amplitude of  the strong pumping laser.  
Numerical simulations of the master equation show that the optimal antibunching scales as $1/P$, as seen in Fig.~\hyperref[fig:fig2]{2(a)} and in the full contour plots of $g^{(2)}(0)$ versus $\Delta_b$ and $\zeta$ shown in Fig.~\hyperref[fig:fig3]{3} 
The region of the parameter space for optimal performance is  roughly given by $\Delta_{\bar{b}}/\kappa > P^2$ and $1/\sqrt{P}<\zeta<1$. 
These results demonstrate that near the instability, the figure of merit for observing the photon blockade is $P\gg 1$ and not simply $g_0/\kappa \gg 1$.

There is an additional constraint that, in order to use the resonant $\bar{b}^\dagger a d$ interaction term, the photon tunneling rate $J$ must be much smaller than the mechanical frequency $\omega_m$.  For the membrane in the middle setup, these conditions may be challenging to achieve due to the high reflectivity required for the membrane.  This could be circumvented by instead utilizing the $a^{\dagger} d d$ nonlinearity, which has no such requirement. One can also consider  using differential modes in `zipper' optomechanical crystals \cite{Chan09}, where making the photon tunneling rate can be tuned over a wide range by controlling the separation between the two cavities. 

Finally, successfully working near the instability requires  the classical power fluctuations in the pump laser to be small enough to prevent the system  from crossing the instability.  More precisely,  the amplitude fluctuations in the pump must be less than the instability parameter $\zeta$ (defined below Eq.\, (\ref{diagH})), which has an optimum value greater than $1/\sqrt{P}$; thus, for $P$ less than $10^3$, this only requires stabilizing the pump power below the $5\, \%$ level, which is readily achievable.

{\textit{Case study}} -- Experimentally these effects could be observed for systems with strong sideband resolution $\omega_m \gg \kappa$ and relatively large single photon optomechanical coupling $g_0\sim \kappa$.  Hybrid photonic-phononic crystals are a promising route to realize both these constraints \cite{Chan2011}, as are mechanical membranes placed in the middle of a high-finesse optical cavity as illustrated in Fig.~\hyperref[fig:fig1]{1(a)} \cite{Thompson08}.  State of the art photonic-phononic crystals have achieved optomechanical coupling $g_0/2\pi$ above $1~\mathrm{MHz}$ \cite{Chan12, Gomis-Bresco14}  and mechanical frequency $\omega_m/2\pi \sim 10~\mathrm{GHz}$ \cite{Safavi-Naeini14}. Optical quality factors as high as nine million have also been reported in silicon photonic crystal cavities, which gives cavity decay rate of $\kappa/2\pi \sim 20~\mathrm{MHz}$ \cite{Sekoguchi14}. In such a case with $g_0/\kappa = 0.1$ and $\omega_m/\kappa = 500$,   $P$ can be as large as $5$ in current devices.  Fig.\ \hyperref[fig:fig3]{3(b)} shows the full range of antibunching obtainable for this $P$, in the optimal case we find that it can be as small as $0.8$, more than an order of magnitude improvement compared to what would be expected away from the instability $\sim 0.99$ \cite{Borkje2013}.  To satisfy the condition $\gamma_{\uparrow} \ll \kappa$, we need $\omega_m/4\zeta\Delta_b \ll 1$ and $\gamma_m\bar{n}_{\mathrm{th}}/2\zeta\ll \kappa$, which imply $\Delta_b \gg \omega_m/4\zeta$ and $\bar{n}_{\mathrm{th}} \ll 2\zeta\kappa/\gamma_m$. This gives an minimum requirement on the $\mathrm{Q}\cdot \mathrm{frequency}$ product:  $Q_m\cdot \omega_m/2\pi > \omega_m/2\zeta\kappa \cdot k_B T/h$. While this case study works in the cryogenic regime, in principle, room temperature operation  may be possible for mechanical oscillators at frequency above $10~\mathrm{GHz}$ and quality factors above $10^6$.

In conclusion, we have presented a scheme to realize few-photon interactions in strongly driven, two-mode optomechanical systems.   Our approach establishes a new figure of merit for realizing strong optomechanical coupling and demonstrates that current  devices, previously thought to have weak coupling,  can  actually be pushed into the regime of strong single-photon nonlinearity.  This would allow one to achieve deterministic entanglement of light in optomechanical systems, which has far-ranging applications in quantum information science.  

We thank A. Clerk, O. Painter, M. Hafezi, J. Lawall, K. Sinha and K. Srinivasan for helpful discussions.  Funding is provided by DARPA QuASAR and the NSF Physics Frontier at the JQI. 

\nocite{*}

\bibliography{SingPhtoNLOM_v15_arxiv}

\providecommand{\noopsort}[1]{}\providecommand{\singleletter}[1]{#1}%
\begin{thebibliography}{48}%
\makeatletter
\providecommand \@ifxundefined [1]{%
 \@ifx{#1\undefined}
}%
\providecommand \@ifnum [1]{%
 \ifnum #1\expandafter \@firstoftwo
 \else \expandafter \@secondoftwo
 \fi
}%
\providecommand \@ifx [1]{%
 \ifx #1\expandafter \@firstoftwo
 \else \expandafter \@secondoftwo
 \fi
}%
\providecommand \natexlab [1]{#1}%
\providecommand \enquote  [1]{``#1''}%
\providecommand \bibnamefont  [1]{#1}%
\providecommand \bibfnamefont [1]{#1}%
\providecommand \citenamefont [1]{#1}%
\providecommand \href@noop [0]{\@secondoftwo}%
\providecommand \href [0]{\begingroup \@sanitize@url \@href}%
\providecommand \@href[1]{\@@startlink{#1}\@@href}%
\providecommand \@@href[1]{\endgroup#1\@@endlink}%
\providecommand \@sanitize@url [0]{\catcode `\\12\catcode `\$12\catcode
  `\&12\catcode `\#12\catcode `\^12\catcode `\_12\catcode `\%12\relax}%
\providecommand \@@startlink[1]{}%
\providecommand \@@endlink[0]{}%
\providecommand \url  [0]{\begingroup\@sanitize@url \@url }%
\providecommand \@url [1]{\endgroup\@href {#1}{\urlprefix }}%
\providecommand \urlprefix  [0]{URL }%
\providecommand \Eprint [0]{\href }%
\providecommand \doibase [0]{http://dx.doi.org/}%
\providecommand \selectlanguage [0]{\@gobble}%
\providecommand \bibinfo  [0]{\@secondoftwo}%
\providecommand \bibfield  [0]{\@secondoftwo}%
\providecommand \translation [1]{[#1]}%
\providecommand \BibitemOpen [0]{}%
\providecommand \bibitemStop [0]{}%
\providecommand \bibitemNoStop [0]{.\EOS\space}%
\providecommand \EOS [0]{\spacefactor3000\relax}%
\providecommand \BibitemShut  [1]{\csname bibitem#1\endcsname}%
\let\auto@bib@innerbib\@empty
\bibitem [{\citenamefont {Milburn}(1989)}]{Milburn89}%
  \BibitemOpen
  \bibfield  {author} {\bibinfo {author} {\bibfnamefont {G.~J.}\ \bibnamefont
  {Milburn}},\ }\href {\doibase 10.1103/PhysRevLett.62.2124} {\bibfield
  {journal} {\bibinfo  {journal} {Phys. Rev. Lett.}\ }\textbf {\bibinfo
  {volume} {62}},\ \bibinfo {pages} {2124} (\bibinfo {year}
  {1989})}\BibitemShut {NoStop}%
\bibitem [{\citenamefont {Turchette}\ \emph {et~al.}(1995)\citenamefont
  {Turchette}, \citenamefont {Hood}, \citenamefont {Lange}, \citenamefont
  {Mabuchi},\ and\ \citenamefont {Kimble}}]{Turchette95}%
  \BibitemOpen
  \bibfield  {author} {\bibinfo {author} {\bibfnamefont {Q.~A.}\ \bibnamefont
  {Turchette}}, \bibinfo {author} {\bibfnamefont {C.~J.}\ \bibnamefont {Hood}},
  \bibinfo {author} {\bibfnamefont {W.}~\bibnamefont {Lange}}, \bibinfo
  {author} {\bibfnamefont {H.}~\bibnamefont {Mabuchi}}, \ and\ \bibinfo
  {author} {\bibfnamefont {H.~J.}\ \bibnamefont {Kimble}},\ }\href {\doibase
  10.1103/PhysRevLett.75.4710} {\bibfield  {journal} {\bibinfo  {journal}
  {Phys. Rev. Lett.}\ }\textbf {\bibinfo {volume} {75}},\ \bibinfo {pages}
  {4710} (\bibinfo {year} {1995})}\BibitemShut {NoStop}%
\bibitem [{\citenamefont {Caulfield}\ and\ \citenamefont
  {Dolev}(2010)}]{Caulfield10}%
  \BibitemOpen
  \bibfield  {author} {\bibinfo {author} {\bibfnamefont {H.~J.}\ \bibnamefont
  {Caulfield}}\ and\ \bibinfo {author} {\bibfnamefont {S.}~\bibnamefont
  {Dolev}},\ }\href {\doibase 10.1038/nphoton.2010.94} {\bibfield  {journal}
  {\bibinfo  {journal} {Nature Photon.}\ }\textbf {\bibinfo {volume} {4}},\
  \bibinfo {pages} {261} (\bibinfo {year} {2010})}\BibitemShut {NoStop}%
\bibitem [{\citenamefont {O'Brien}\ \emph {et~al.}(2009)\citenamefont
  {O'Brien}, \citenamefont {Furusawa},\ and\ \citenamefont
  {Vu\v{c}kovi\'{c}}}]{Obrien09}%
  \BibitemOpen
  \bibfield  {author} {\bibinfo {author} {\bibfnamefont {J.~L.}\ \bibnamefont
  {O'Brien}}, \bibinfo {author} {\bibfnamefont {A.}~\bibnamefont {Furusawa}}, \
  and\ \bibinfo {author} {\bibfnamefont {J.}~\bibnamefont {Vu\v{c}kovi\'{c}}},\
  }\href {\doibase 10.1038/nphoton.2009.229} {\bibfield  {journal} {\bibinfo
  {journal} {Nature Photon.}\ }\textbf {\bibinfo {volume} {3}},\ \bibinfo
  {pages} {687} (\bibinfo {year} {2009})}\BibitemShut {NoStop}%
\bibitem [{\citenamefont {Birnbaum}\ \emph {et~al.}(2005)\citenamefont
  {Birnbaum}, \citenamefont {Boca}, \citenamefont {Miller}, \citenamefont
  {Boozer}, \citenamefont {Northup},\ and\ \citenamefont
  {Kimble}}]{Birnbaum2005}%
  \BibitemOpen
  \bibfield  {author} {\bibinfo {author} {\bibfnamefont {K.~M.}\ \bibnamefont
  {Birnbaum}}, \bibinfo {author} {\bibfnamefont {A.}~\bibnamefont {Boca}},
  \bibinfo {author} {\bibfnamefont {R.}~\bibnamefont {Miller}}, \bibinfo
  {author} {\bibfnamefont {A.~D.}\ \bibnamefont {Boozer}}, \bibinfo {author}
  {\bibfnamefont {T.~E.}\ \bibnamefont {Northup}}, \ and\ \bibinfo {author}
  {\bibfnamefont {H.~J.}\ \bibnamefont {Kimble}},\ }\href {\doibase
  10.1038/nature03804} {\bibfield  {journal} {\bibinfo  {journal} {Nature}\
  }\textbf {\bibinfo {volume} {436}},\ \bibinfo {pages} {87} (\bibinfo {year}
  {2005})}\BibitemShut {NoStop}%
\bibitem [{\citenamefont {Press}\ \emph {et~al.}(2007)\citenamefont {Press},
  \citenamefont {G\"otzinger}, \citenamefont {Reitzenstein}, \citenamefont
  {Hofmann}, \citenamefont {L\"offler}, \citenamefont {Kamp}, \citenamefont
  {Forchel},\ and\ \citenamefont {Yamamoto}}]{Press07}%
  \BibitemOpen
  \bibfield  {author} {\bibinfo {author} {\bibfnamefont {D.}~\bibnamefont
  {Press}}, \bibinfo {author} {\bibfnamefont {S.}~\bibnamefont {G\"otzinger}},
  \bibinfo {author} {\bibfnamefont {S.}~\bibnamefont {Reitzenstein}}, \bibinfo
  {author} {\bibfnamefont {C.}~\bibnamefont {Hofmann}}, \bibinfo {author}
  {\bibfnamefont {A.}~\bibnamefont {L\"offler}}, \bibinfo {author}
  {\bibfnamefont {M.}~\bibnamefont {Kamp}}, \bibinfo {author} {\bibfnamefont
  {A.}~\bibnamefont {Forchel}}, \ and\ \bibinfo {author} {\bibfnamefont
  {Y.}~\bibnamefont {Yamamoto}},\ }\href {\doibase
  10.1103/PhysRevLett.98.117402} {\bibfield  {journal} {\bibinfo  {journal}
  {Phys. Rev. Lett.}\ }\textbf {\bibinfo {volume} {98}},\ \bibinfo {pages}
  {117402} (\bibinfo {year} {2007})}\BibitemShut {NoStop}%
\bibitem [{\citenamefont {Bishop}\ \emph {et~al.}(2009)\citenamefont {Bishop},
  \citenamefont {Chow}, \citenamefont {Koch}, \citenamefont {Houck},
  \citenamefont {Devoret}, \citenamefont {Thuneberg}, \citenamefont {Girvin},\
  and\ \citenamefont {Schoelkopf}}]{Bishop2009}%
  \BibitemOpen
  \bibfield  {author} {\bibinfo {author} {\bibfnamefont {L.~S.}\ \bibnamefont
  {Bishop}}, \bibinfo {author} {\bibfnamefont {J.~M.}\ \bibnamefont {Chow}},
  \bibinfo {author} {\bibfnamefont {J.}~\bibnamefont {Koch}}, \bibinfo {author}
  {\bibfnamefont {A.~A.}\ \bibnamefont {Houck}}, \bibinfo {author}
  {\bibfnamefont {M.~H.}\ \bibnamefont {Devoret}}, \bibinfo {author}
  {\bibfnamefont {E.}~\bibnamefont {Thuneberg}}, \bibinfo {author}
  {\bibfnamefont {S.~M.}\ \bibnamefont {Girvin}}, \ and\ \bibinfo {author}
  {\bibfnamefont {R.~J.}\ \bibnamefont {Schoelkopf}},\ }\href {\doibase
  10.1038/nphys1154} {\bibfield  {journal} {\bibinfo  {journal} {Nature Phys.}\
  }\textbf {\bibinfo {volume} {5}},\ \bibinfo {pages} {105} (\bibinfo {year}
  {2009})}\BibitemShut {NoStop}%
\bibitem [{\citenamefont {Volz}\ \emph {et~al.}(2012)\citenamefont {Volz},
  \citenamefont {Reinhard}, \citenamefont {Winger}, \citenamefont {Badolato},
  \citenamefont {Hennessy},\ and\ \citenamefont {Hu}}]{Volz2012}%
  \BibitemOpen
  \bibfield  {author} {\bibinfo {author} {\bibfnamefont {T.}~\bibnamefont
  {Volz}}, \bibinfo {author} {\bibfnamefont {A.}~\bibnamefont {Reinhard}},
  \bibinfo {author} {\bibfnamefont {M.}~\bibnamefont {Winger}}, \bibinfo
  {author} {\bibfnamefont {A.}~\bibnamefont {Badolato}}, \bibinfo {author}
  {\bibfnamefont {K.~J.}\ \bibnamefont {Hennessy}}, \ and\ \bibinfo {author}
  {\bibfnamefont {E.~L.}\ \bibnamefont {Hu}},\ }\href {\doibase
  10.1038/NPHOTON.2012.181} {\bibfield  {journal} {\bibinfo  {journal} {Nature
  Photon.}\ }\textbf {\bibinfo {volume} {6}},\ \bibinfo {pages} {605} (\bibinfo
  {year} {2012})}\BibitemShut {NoStop}%
\bibitem [{\citenamefont {Bose}\ \emph {et~al.}(2012)\citenamefont {Bose},
  \citenamefont {Sridharan}, \citenamefont {Kim}, \citenamefont {Solomon},\
  and\ \citenamefont {Waks}}]{Bose12}%
  \BibitemOpen
  \bibfield  {author} {\bibinfo {author} {\bibfnamefont {R.}~\bibnamefont
  {Bose}}, \bibinfo {author} {\bibfnamefont {D.}~\bibnamefont {Sridharan}},
  \bibinfo {author} {\bibfnamefont {H.}~\bibnamefont {Kim}}, \bibinfo {author}
  {\bibfnamefont {G.~S.}\ \bibnamefont {Solomon}}, \ and\ \bibinfo {author}
  {\bibfnamefont {E.}~\bibnamefont {Waks}},\ }\href {\doibase
  10.1103/PhysRevLett.108.227402} {\bibfield  {journal} {\bibinfo  {journal}
  {Phys. Rev. Lett.}\ }\textbf {\bibinfo {volume} {108}},\ \bibinfo {pages}
  {227402} (\bibinfo {year} {2012})}\BibitemShut {NoStop}%
\bibitem [{\citenamefont {Chen}\ \emph {et~al.}(2013)\citenamefont {Chen},
  \citenamefont {Beck}, \citenamefont {B\"{u}cker}, \citenamefont {Gullans},
  \citenamefont {Lukin}, \citenamefont {Tanji-Suzuki},\ and\ \citenamefont
  {Vuleti\'{c}}}]{Chen13}%
  \BibitemOpen
  \bibfield  {author} {\bibinfo {author} {\bibfnamefont {W.}~\bibnamefont
  {Chen}}, \bibinfo {author} {\bibfnamefont {K.~M.}\ \bibnamefont {Beck}},
  \bibinfo {author} {\bibfnamefont {R.}~\bibnamefont {B\"{u}cker}}, \bibinfo
  {author} {\bibfnamefont {M.}~\bibnamefont {Gullans}}, \bibinfo {author}
  {\bibfnamefont {M.~D.}\ \bibnamefont {Lukin}}, \bibinfo {author}
  {\bibfnamefont {H.}~\bibnamefont {Tanji-Suzuki}}, \ and\ \bibinfo {author}
  {\bibfnamefont {V.}~\bibnamefont {Vuleti\'{c}}},\ }\href {\doibase
  10.1126/science.1238169} {\bibfield  {journal} {\bibinfo  {journal}
  {Science}\ }\textbf {\bibinfo {volume} {341}},\ \bibinfo {pages} {768}
  (\bibinfo {year} {2013})}\BibitemShut {NoStop}%
\bibitem [{\citenamefont {Reiserer}\ \emph {et~al.}(2014)\citenamefont
  {Reiserer}, \citenamefont {Kalb}, \citenamefont {Rempe},\ and\ \citenamefont
  {Ritter}}]{Reiserer14}%
  \BibitemOpen
  \bibfield  {author} {\bibinfo {author} {\bibfnamefont {A.}~\bibnamefont
  {Reiserer}}, \bibinfo {author} {\bibfnamefont {N.}~\bibnamefont {Kalb}},
  \bibinfo {author} {\bibfnamefont {G.}~\bibnamefont {Rempe}}, \ and\ \bibinfo
  {author} {\bibfnamefont {S.}~\bibnamefont {Ritter}},\ }\href
  {http://www.nature.com/nature/journal/v508/n7495/full/nature13177.html}
  {\bibfield  {journal} {\bibinfo  {journal} {Nature}\ }\textbf {\bibinfo
  {volume} {508}},\ \bibinfo {pages} {237} (\bibinfo {year}
  {2014})}\BibitemShut {NoStop}%
\bibitem [{\citenamefont {Tiecke}\ \emph {et~al.}(2014)\citenamefont {Tiecke},
  \citenamefont {Thompson}, \citenamefont {de~Leon}, \citenamefont {Liu},
  \citenamefont {Vuletic},\ and\ \citenamefont {Lukin}}]{Tiecke14}%
  \BibitemOpen
  \bibfield  {author} {\bibinfo {author} {\bibfnamefont {T.~G.}\ \bibnamefont
  {Tiecke}}, \bibinfo {author} {\bibfnamefont {J.~D.}\ \bibnamefont
  {Thompson}}, \bibinfo {author} {\bibfnamefont {N.~P.}\ \bibnamefont
  {de~Leon}}, \bibinfo {author} {\bibfnamefont {L.~R.}\ \bibnamefont {Liu}},
  \bibinfo {author} {\bibfnamefont {V.}~\bibnamefont {Vuletic}}, \ and\
  \bibinfo {author} {\bibfnamefont {M.~D.}\ \bibnamefont {Lukin}},\ }\href
  {http://www.nature.com/nature/journal/v508/n7495/full/nature13188.html}
  {\bibfield  {journal} {\bibinfo  {journal} {Nature}\ }\textbf {\bibinfo
  {volume} {508}},\ \bibinfo {pages} {241} (\bibinfo {year}
  {2014})}\BibitemShut {NoStop}%
\bibitem [{\citenamefont {Imamoglu}\ \emph {et~al.}(1997)\citenamefont
  {Imamoglu}, \citenamefont {Schmidt}, \citenamefont {Woods},\ and\
  \citenamefont {Deutsch}}]{Imamoglu1997}%
  \BibitemOpen
  \bibfield  {author} {\bibinfo {author} {\bibfnamefont {A.}~\bibnamefont
  {Imamoglu}}, \bibinfo {author} {\bibfnamefont {H.}~\bibnamefont {Schmidt}},
  \bibinfo {author} {\bibfnamefont {G.}~\bibnamefont {Woods}}, \ and\ \bibinfo
  {author} {\bibfnamefont {M.}~\bibnamefont {Deutsch}},\ }\href {\doibase
  10.1103/PhysRevLett.79.1467} {\bibfield  {journal} {\bibinfo  {journal}
  {Phys. Rev. Lett.}\ }\textbf {\bibinfo {volume} {79}},\ \bibinfo {pages}
  {1467} (\bibinfo {year} {1997})}\BibitemShut {NoStop}%
\bibitem [{\citenamefont {Lukin}\ and\ \citenamefont
  {Imamoglu}(2000)}]{Lukin00}%
  \BibitemOpen
  \bibfield  {author} {\bibinfo {author} {\bibfnamefont {M.~D.}\ \bibnamefont
  {Lukin}}\ and\ \bibinfo {author} {\bibfnamefont {A.}~\bibnamefont
  {Imamoglu}},\ }\href {\doibase 10.1103/PhysRevLett.84.1419} {\bibfield
  {journal} {\bibinfo  {journal} {Phys. Rev. Lett.}\ }\textbf {\bibinfo
  {volume} {84}},\ \bibinfo {pages} {1419} (\bibinfo {year}
  {2000})}\BibitemShut {NoStop}%
\bibitem [{\citenamefont {Lin}\ \emph {et~al.}(2012)\citenamefont {Lin},
  \citenamefont {Wu}, \citenamefont {Shiau}, \citenamefont {Chen},
  \citenamefont {Yu}, \citenamefont {Chen},\ and\ \citenamefont
  {Chen}}]{Lin12}%
  \BibitemOpen
  \bibfield  {author} {\bibinfo {author} {\bibfnamefont {C.-C.}\ \bibnamefont
  {Lin}}, \bibinfo {author} {\bibfnamefont {M.-C.}\ \bibnamefont {Wu}},
  \bibinfo {author} {\bibfnamefont {B.-W.}\ \bibnamefont {Shiau}}, \bibinfo
  {author} {\bibfnamefont {Y.-H.}\ \bibnamefont {Chen}}, \bibinfo {author}
  {\bibfnamefont {I.~A.}\ \bibnamefont {Yu}}, \bibinfo {author} {\bibfnamefont
  {Y.-F.}\ \bibnamefont {Chen}}, \ and\ \bibinfo {author} {\bibfnamefont
  {Y.-C.}\ \bibnamefont {Chen}},\ }\href {\doibase 10.1103/PhysRevA.86.063836}
  {\bibfield  {journal} {\bibinfo  {journal} {Phys. Rev. A}\ }\textbf {\bibinfo
  {volume} {86}},\ \bibinfo {pages} {063836} (\bibinfo {year}
  {2012})}\BibitemShut {NoStop}%
\bibitem [{\citenamefont {Hwang}\ \emph {et~al.}(2009)\citenamefont {Hwang},
  \citenamefont {Pototschnig}, \citenamefont {Lettow}, \citenamefont {Zumofen},
  \citenamefont {Renn}, \citenamefont {G\"{o}tzinger},\ and\ \citenamefont
  {Sandoghdar}}]{Hwang09}%
  \BibitemOpen
  \bibfield  {author} {\bibinfo {author} {\bibfnamefont {J.}~\bibnamefont
  {Hwang}}, \bibinfo {author} {\bibfnamefont {M.}~\bibnamefont {Pototschnig}},
  \bibinfo {author} {\bibfnamefont {R.}~\bibnamefont {Lettow}}, \bibinfo
  {author} {\bibfnamefont {G.}~\bibnamefont {Zumofen}}, \bibinfo {author}
  {\bibfnamefont {A.}~\bibnamefont {Renn}}, \bibinfo {author} {\bibfnamefont
  {S.}~\bibnamefont {G\"{o}tzinger}}, \ and\ \bibinfo {author} {\bibfnamefont
  {V.}~\bibnamefont {Sandoghdar}},\ }\href {\doibase 10.1038/nature08134}
  {\bibfield  {journal} {\bibinfo  {journal} {Nature}\ }\textbf {\bibinfo
  {volume} {460}},\ \bibinfo {pages} {76} (\bibinfo {year} {2009})}\BibitemShut
  {NoStop}%
\bibitem [{\citenamefont {Peyronel}\ \emph {et~al.}(2012)\citenamefont
  {Peyronel}, \citenamefont {Firstenberg}, \citenamefont {Liang}, \citenamefont
  {Hofferberth}, \citenamefont {Gorshkov}, \citenamefont {Pohl}, \citenamefont
  {Lukin},\ and\ \citenamefont {Vuleti\'{c}}}]{Peyronel2012}%
  \BibitemOpen
  \bibfield  {author} {\bibinfo {author} {\bibfnamefont {T.}~\bibnamefont
  {Peyronel}}, \bibinfo {author} {\bibfnamefont {O.}~\bibnamefont
  {Firstenberg}}, \bibinfo {author} {\bibfnamefont {Q.-Y.}\ \bibnamefont
  {Liang}}, \bibinfo {author} {\bibfnamefont {S.}~\bibnamefont {Hofferberth}},
  \bibinfo {author} {\bibfnamefont {A.~V.}\ \bibnamefont {Gorshkov}}, \bibinfo
  {author} {\bibfnamefont {T.}~\bibnamefont {Pohl}}, \bibinfo {author}
  {\bibfnamefont {M.~D.}\ \bibnamefont {Lukin}}, \ and\ \bibinfo {author}
  {\bibfnamefont {V.}~\bibnamefont {Vuleti\'{c}}},\ }\href {\doibase
  10.1038/nature11361} {\bibfield  {journal} {\bibinfo  {journal} {Nature}\
  }\textbf {\bibinfo {volume} {488}},\ \bibinfo {pages} {57} (\bibinfo {year}
  {2012})}\BibitemShut {NoStop}%
\bibitem [{\citenamefont {Dudin}\ and\ \citenamefont
  {Kuzmich}(2012)}]{Dudin12}%
  \BibitemOpen
  \bibfield  {author} {\bibinfo {author} {\bibfnamefont {Y.~O.}\ \bibnamefont
  {Dudin}}\ and\ \bibinfo {author} {\bibfnamefont {A.}~\bibnamefont
  {Kuzmich}},\ }\href {\doibase 10.1126/science.1217901} {\bibfield  {journal}
  {\bibinfo  {journal} {Science}\ }\textbf {\bibinfo {volume} {336}},\ \bibinfo
  {pages} {887} (\bibinfo {year} {2012})}\BibitemShut {NoStop}%
\bibitem [{\citenamefont {Gorniaczyk}\ \emph {et~al.}(2014)\citenamefont
  {Gorniaczyk}, \citenamefont {C.~Tresp}, \citenamefont {Fedder},\ and\
  \citenamefont {Hofferberth}}]{Gorn14}%
  \BibitemOpen
  \bibfield  {author} {\bibinfo {author} {\bibfnamefont {H.}~\bibnamefont
  {Gorniaczyk}}, \bibinfo {author} {\bibfnamefont {J.~S.}\ \bibnamefont
  {C.~Tresp}}, \bibinfo {author} {\bibfnamefont {H.}~\bibnamefont {Fedder}}, \
  and\ \bibinfo {author} {\bibfnamefont {S.}~\bibnamefont {Hofferberth}},\
  }\href {http://arxiv.org/abs/1404.2876} {} (\bibinfo {year} {2014}),\ \Eprint
  {http://arxiv.org/abs/arXiv:1404.2876} {arXiv:1404.2876} \BibitemShut
  {NoStop}%
\bibitem [{\citenamefont {Tiarks}\ \emph {et~al.}(2014)\citenamefont {Tiarks},
  \citenamefont {Baur}, \citenamefont {Schneider}, \citenamefont {D{\"u}rr},\
  and\ \citenamefont {Rempe}}]{Tiarks14}%
  \BibitemOpen
  \bibfield  {author} {\bibinfo {author} {\bibfnamefont {D.}~\bibnamefont
  {Tiarks}}, \bibinfo {author} {\bibfnamefont {S.}~\bibnamefont {Baur}},
  \bibinfo {author} {\bibfnamefont {K.}~\bibnamefont {Schneider}}, \bibinfo
  {author} {\bibfnamefont {S.}~\bibnamefont {D{\"u}rr}}, \ and\ \bibinfo
  {author} {\bibfnamefont {G.}~\bibnamefont {Rempe}},\ }\href
  {http://arxiv-web3.library.cornell.edu/abs/1404.3061v1} {} (\bibinfo {year}
  {2014}),\ \Eprint {http://arxiv.org/abs/arXiv:1404.3061} {arXiv:1404.3061}
  \BibitemShut {NoStop}%
\bibitem [{\citenamefont {Chang}\ \emph {et~al.}(2007)\citenamefont {Chang},
  \citenamefont {S{\o}rensen}, \citenamefont {Demler},\ and\ \citenamefont
  {Lukin}}]{Chang07}%
  \BibitemOpen
  \bibfield  {author} {\bibinfo {author} {\bibfnamefont {D.~E.}\ \bibnamefont
  {Chang}}, \bibinfo {author} {\bibfnamefont {A.~S.}\ \bibnamefont
  {S{\o}rensen}}, \bibinfo {author} {\bibfnamefont {E.~A.}\ \bibnamefont
  {Demler}}, \ and\ \bibinfo {author} {\bibfnamefont {M.~D.}\ \bibnamefont
  {Lukin}},\ }\href {\doibase 10.1038/nphys708} {\bibfield  {journal} {\bibinfo
   {journal} {Nature Phys.}\ }\textbf {\bibinfo {volume} {3}},\ \bibinfo
  {pages} {807} (\bibinfo {year} {2007})}\BibitemShut {NoStop}%
\bibitem [{\citenamefont {Gullans}\ \emph {et~al.}(2013)\citenamefont
  {Gullans}, \citenamefont {Chang}, \citenamefont {Koppens}, \citenamefont
  {Garc{\' \i}a~de Abajo},\ and\ \citenamefont {Lukin}}]{Gullans2013}%
  \BibitemOpen
  \bibfield  {author} {\bibinfo {author} {\bibfnamefont {M.}~\bibnamefont
  {Gullans}}, \bibinfo {author} {\bibfnamefont {D.~E.}\ \bibnamefont {Chang}},
  \bibinfo {author} {\bibfnamefont {F.~H.~L.}\ \bibnamefont {Koppens}},
  \bibinfo {author} {\bibfnamefont {F.~J.}\ \bibnamefont {Garc{\' \i}a~de
  Abajo}}, \ and\ \bibinfo {author} {\bibfnamefont {M.~D.}\ \bibnamefont
  {Lukin}},\ }\href {\doibase 10.1103/PhysRevLett.111.247401} {\bibfield
  {journal} {\bibinfo  {journal} {Phys. Rev. Lett.}\ }\textbf {\bibinfo
  {volume} {111}},\ \bibinfo {pages} {247401} (\bibinfo {year}
  {2013})}\BibitemShut {NoStop}%
\bibitem [{\citenamefont {Bajcsy}\ \emph {et~al.}(2009)\citenamefont {Bajcsy},
  \citenamefont {Hofferberth}, \citenamefont {Balic}, \citenamefont {Peyronel},
  \citenamefont {Hafezi}, \citenamefont {Zibrov}, \citenamefont {Vuletic},\
  and\ \citenamefont {Lukin}}]{Bajcsy09}%
  \BibitemOpen
  \bibfield  {author} {\bibinfo {author} {\bibfnamefont {M.}~\bibnamefont
  {Bajcsy}}, \bibinfo {author} {\bibfnamefont {S.}~\bibnamefont {Hofferberth}},
  \bibinfo {author} {\bibfnamefont {V.}~\bibnamefont {Balic}}, \bibinfo
  {author} {\bibfnamefont {T.}~\bibnamefont {Peyronel}}, \bibinfo {author}
  {\bibfnamefont {M.}~\bibnamefont {Hafezi}}, \bibinfo {author} {\bibfnamefont
  {A.~S.}\ \bibnamefont {Zibrov}}, \bibinfo {author} {\bibfnamefont
  {V.}~\bibnamefont {Vuletic}}, \ and\ \bibinfo {author} {\bibfnamefont
  {M.~D.}\ \bibnamefont {Lukin}},\ }\href {\doibase
  10.1103/PhysRevLett.102.203902} {\bibfield  {journal} {\bibinfo  {journal}
  {Phys. Rev. Lett.}\ }\textbf {\bibinfo {volume} {102}},\ \bibinfo {pages}
  {203902} (\bibinfo {year} {2009})}\BibitemShut {NoStop}%
\bibitem [{\citenamefont {Venkataraman}\ \emph {et~al.}(2011)\citenamefont
  {Venkataraman}, \citenamefont {Saha}, \citenamefont {Londero},\ and\
  \citenamefont {Gaeta}}]{Venk11}%
  \BibitemOpen
  \bibfield  {author} {\bibinfo {author} {\bibfnamefont {V.}~\bibnamefont
  {Venkataraman}}, \bibinfo {author} {\bibfnamefont {K.}~\bibnamefont {Saha}},
  \bibinfo {author} {\bibfnamefont {P.}~\bibnamefont {Londero}}, \ and\
  \bibinfo {author} {\bibfnamefont {A.~L.}\ \bibnamefont {Gaeta}},\ }\href
  {\doibase 10.1103/PhysRevLett.107.193902} {\bibfield  {journal} {\bibinfo
  {journal} {Phys. Rev. Lett.}\ }\textbf {\bibinfo {volume} {107}},\ \bibinfo
  {pages} {193902} (\bibinfo {year} {2011})}\BibitemShut {NoStop}%
\bibitem [{\citenamefont {Kolchin}\ \emph {et~al.}(2011)\citenamefont
  {Kolchin}, \citenamefont {Oulton},\ and\ \citenamefont
  {Zhang}}]{Kolchin2011}%
  \BibitemOpen
  \bibfield  {author} {\bibinfo {author} {\bibfnamefont {P.}~\bibnamefont
  {Kolchin}}, \bibinfo {author} {\bibfnamefont {R.~F.}\ \bibnamefont {Oulton}},
  \ and\ \bibinfo {author} {\bibfnamefont {X.}~\bibnamefont {Zhang}},\ }\href
  {\doibase 10.1103/PhysRevLett.106.113601} {\bibfield  {journal} {\bibinfo
  {journal} {Phys. Rev. Lett.}\ }\textbf {\bibinfo {volume} {106}},\ \bibinfo
  {pages} {113601} (\bibinfo {year} {2011})}\BibitemShut {NoStop}%
\bibitem [{\citenamefont {O'Shea}\ \emph {et~al.}(2013)\citenamefont {O'Shea},
  \citenamefont {Junge}, \citenamefont {Volz},\ and\ \citenamefont
  {Rauschenbeutel}}]{OShea13}%
  \BibitemOpen
  \bibfield  {author} {\bibinfo {author} {\bibfnamefont {D.}~\bibnamefont
  {O'Shea}}, \bibinfo {author} {\bibfnamefont {C.}~\bibnamefont {Junge}},
  \bibinfo {author} {\bibfnamefont {J.}~\bibnamefont {Volz}}, \ and\ \bibinfo
  {author} {\bibfnamefont {A.}~\bibnamefont {Rauschenbeutel}},\ }\href
  {\doibase 10.1103/PhysRevLett.111.193601} {\bibfield  {journal} {\bibinfo
  {journal} {Phys. Rev. Lett.}\ }\textbf {\bibinfo {volume} {111}},\ \bibinfo
  {pages} {193601} (\bibinfo {year} {2013})}\BibitemShut {NoStop}%
\bibitem [{\citenamefont {Kippenberg}\ and\ \citenamefont
  {Vahala}(2008)}]{Kippenberg2008}%
  \BibitemOpen
  \bibfield  {author} {\bibinfo {author} {\bibfnamefont {T.~J.}\ \bibnamefont
  {Kippenberg}}\ and\ \bibinfo {author} {\bibfnamefont {K.~J.}\ \bibnamefont
  {Vahala}},\ }\href {http://www.sciencemag.org/content/321/5893/1172.full}
  {\bibfield  {journal} {\bibinfo  {journal} {Science}\ }\textbf {\bibinfo
  {volume} {321}},\ \bibinfo {pages} {1172} (\bibinfo {year}
  {2008})}\BibitemShut {NoStop}%
\bibitem [{\citenamefont {Aspelmeyer}\ \emph {et~al.}(2013)\citenamefont
  {Aspelmeyer}, \citenamefont {Kippenberg},\ and\ \citenamefont
  {Marquardt}}]{Aspelmeyer2013}%
  \BibitemOpen
  \bibfield  {author} {\bibinfo {author} {\bibfnamefont {M.}~\bibnamefont
  {Aspelmeyer}}, \bibinfo {author} {\bibfnamefont {T.~J.}\ \bibnamefont
  {Kippenberg}}, \ and\ \bibinfo {author} {\bibfnamefont {F.}~\bibnamefont
  {Marquardt}},\ }\href {http://arxiv.org/abs/1303.0733} {} (\bibinfo {year}
  {2013}),\ \Eprint {http://arxiv.org/abs/arXiv:1303.0733} {arXiv:1303.0733}
  \BibitemShut {NoStop}%
\bibitem [{\citenamefont {Thompson}\ \emph {et~al.}(2008)\citenamefont
  {Thompson}, \citenamefont {Zwickl}, \citenamefont {Jayich}, \citenamefont
  {Marquardt}, \citenamefont {Girvin},\ and\ \citenamefont
  {Harris}}]{Thompson08}%
  \BibitemOpen
  \bibfield  {author} {\bibinfo {author} {\bibfnamefont {J.~D.}\ \bibnamefont
  {Thompson}}, \bibinfo {author} {\bibfnamefont {B.~M.}\ \bibnamefont
  {Zwickl}}, \bibinfo {author} {\bibfnamefont {A.~M.}\ \bibnamefont {Jayich}},
  \bibinfo {author} {\bibfnamefont {F.}~\bibnamefont {Marquardt}}, \bibinfo
  {author} {\bibfnamefont {S.~M.}\ \bibnamefont {Girvin}}, \ and\ \bibinfo
  {author} {\bibfnamefont {J.~G.~E.}\ \bibnamefont {Harris}},\ }\href
  {http://www.nature.com/nature/journal/v452/n7183/abs/nature06715.html}
  {\bibfield  {journal} {\bibinfo  {journal} {Nature}\ }\textbf {\bibinfo
  {volume} {452}},\ \bibinfo {pages} {72} (\bibinfo {year} {2008})}\BibitemShut
  {NoStop}%
\bibitem [{\citenamefont {Chan}\ \emph {et~al.}(2011)\citenamefont {Chan},
  \citenamefont {Alegre}, \citenamefont {Safavi-Naeini}, \citenamefont {Hill},
  \citenamefont {Krause}, \citenamefont {Gr\"{o}blacher}, \citenamefont
  {Aspelmeyer},\ and\ \citenamefont {Painter}}]{Chan2011}%
  \BibitemOpen
  \bibfield  {author} {\bibinfo {author} {\bibfnamefont {J.}~\bibnamefont
  {Chan}}, \bibinfo {author} {\bibfnamefont {T.~P.~M.}\ \bibnamefont {Alegre}},
  \bibinfo {author} {\bibfnamefont {A.~H.}\ \bibnamefont {Safavi-Naeini}},
  \bibinfo {author} {\bibfnamefont {J.~T.}\ \bibnamefont {Hill}}, \bibinfo
  {author} {\bibfnamefont {A.}~\bibnamefont {Krause}}, \bibinfo {author}
  {\bibfnamefont {S.}~\bibnamefont {Gr\"{o}blacher}}, \bibinfo {author}
  {\bibfnamefont {M.}~\bibnamefont {Aspelmeyer}}, \ and\ \bibinfo {author}
  {\bibfnamefont {O.}~\bibnamefont {Painter}},\ }\href {\doibase
  10.1038/nature10461} {\bibfield  {journal} {\bibinfo  {journal} {Nature}\
  }\textbf {\bibinfo {volume} {478}},\ \bibinfo {pages} {89} (\bibinfo {year}
  {2011})}\BibitemShut {NoStop}%
\bibitem [{\citenamefont {Verhagen}\ \emph {et~al.}(2012)\citenamefont
  {Verhagen}, \citenamefont {Del使lise}, \citenamefont {Weis}, \citenamefont
  {Schliesser},\ and\ \citenamefont {Kippenberg}}]{Verhagen2012}%
  \BibitemOpen
  \bibfield  {author} {\bibinfo {author} {\bibfnamefont {E.}~\bibnamefont
  {Verhagen}}, \bibinfo {author} {\bibfnamefont {S.}~\bibnamefont {Del使lise}},
  \bibinfo {author} {\bibfnamefont {S.}~\bibnamefont {Weis}}, \bibinfo {author}
  {\bibfnamefont {A.}~\bibnamefont {Schliesser}}, \ and\ \bibinfo {author}
  {\bibfnamefont {T.~J.}\ \bibnamefont {Kippenberg}},\ }\href {\doibase
  10.1038/nature10787} {\bibfield  {journal} {\bibinfo  {journal} {Nature}\
  }\textbf {\bibinfo {volume} {482}},\ \bibinfo {pages} {63} (\bibinfo {year}
  {2012})}\BibitemShut {NoStop}%
\bibitem [{\citenamefont {Gr\"{o}blacher}\ \emph {et~al.}(2009)\citenamefont
  {Gr\"{o}blacher}, \citenamefont {Hammerer}, \citenamefont {Vanner},\ and\
  \citenamefont {Aspelmeyer}}]{Groblacher2009}%
  \BibitemOpen
  \bibfield  {author} {\bibinfo {author} {\bibfnamefont {S.}~\bibnamefont
  {Gr\"{o}blacher}}, \bibinfo {author} {\bibfnamefont {K.}~\bibnamefont
  {Hammerer}}, \bibinfo {author} {\bibfnamefont {M.~R.}\ \bibnamefont
  {Vanner}}, \ and\ \bibinfo {author} {\bibfnamefont {M.}~\bibnamefont
  {Aspelmeyer}},\ }\href {\doibase 10.1038/nature08171} {\bibfield  {journal}
  {\bibinfo  {journal} {Nature}\ }\textbf {\bibinfo {volume} {460}},\ \bibinfo
  {pages} {724} (\bibinfo {year} {2009})}\BibitemShut {NoStop}%
\bibitem [{\citenamefont {Teufel}\ \emph {et~al.}(2011)\citenamefont {Teufel},
  \citenamefont {Li}, \citenamefont {Allman}, \citenamefont {Cicak},
  \citenamefont {Sirois}, \citenamefont {Whittaker},\ and\ \citenamefont
  {Simmonds}}]{Teufel2011}%
  \BibitemOpen
  \bibfield  {author} {\bibinfo {author} {\bibfnamefont {J.~D.}\ \bibnamefont
  {Teufel}}, \bibinfo {author} {\bibfnamefont {D.}~\bibnamefont {Li}}, \bibinfo
  {author} {\bibfnamefont {M.~S.}\ \bibnamefont {Allman}}, \bibinfo {author}
  {\bibfnamefont {K.}~\bibnamefont {Cicak}}, \bibinfo {author} {\bibfnamefont
  {A.~J.}\ \bibnamefont {Sirois}}, \bibinfo {author} {\bibfnamefont {J.~D.}\
  \bibnamefont {Whittaker}}, \ and\ \bibinfo {author} {\bibfnamefont {R.~W.}\
  \bibnamefont {Simmonds}},\ }\href {\doibase 10.1038/nature09898} {\bibfield
  {journal} {\bibinfo  {journal} {Nature}\ }\textbf {\bibinfo {volume} {471}},\
  \bibinfo {pages} {204} (\bibinfo {year} {2011})}\BibitemShut {NoStop}%
\bibitem [{\citenamefont {Lemonde}\ \emph {et~al.}(2013)\citenamefont
  {Lemonde}, \citenamefont {Didier},\ and\ \citenamefont
  {Clerk}}]{Lemonde2013}%
  \BibitemOpen
  \bibfield  {author} {\bibinfo {author} {\bibfnamefont {M.-A.}\ \bibnamefont
  {Lemonde}}, \bibinfo {author} {\bibfnamefont {N.}~\bibnamefont {Didier}}, \
  and\ \bibinfo {author} {\bibfnamefont {A.~A.}\ \bibnamefont {Clerk}},\ }\href
  {\doibase 10.1103/PhysRevLett.111.053602} {\bibfield  {journal} {\bibinfo
  {journal} {Phys. Rev. Lett.}\ }\textbf {\bibinfo {volume} {111}},\ \bibinfo
  {pages} {053602} (\bibinfo {year} {2013})}\BibitemShut {NoStop}%
\bibitem [{\citenamefont {B\o{}rkje}\ \emph {et~al.}(2013)\citenamefont
  {B\o{}rkje}, \citenamefont {Nunnenkamp}, \citenamefont {Teufel},\ and\
  \citenamefont {Girvin}}]{Borkje2013}%
  \BibitemOpen
  \bibfield  {author} {\bibinfo {author} {\bibfnamefont {K.}~\bibnamefont
  {B\o{}rkje}}, \bibinfo {author} {\bibfnamefont {A.}~\bibnamefont
  {Nunnenkamp}}, \bibinfo {author} {\bibfnamefont {J.~D.}\ \bibnamefont
  {Teufel}}, \ and\ \bibinfo {author} {\bibfnamefont {S.~M.}\ \bibnamefont
  {Girvin}},\ }\href {\doibase 10.1103/PhysRevLett.111.053603} {\bibfield
  {journal} {\bibinfo  {journal} {Phys. Rev. Lett.}\ }\textbf {\bibinfo
  {volume} {111}},\ \bibinfo {pages} {053603} (\bibinfo {year}
  {2013})}\BibitemShut {NoStop}%
\bibitem [{\citenamefont {Kronwald}\ and\ \citenamefont
  {Marquardt}(2013)}]{Kronwald2013}%
  \BibitemOpen
  \bibfield  {author} {\bibinfo {author} {\bibfnamefont {A.}~\bibnamefont
  {Kronwald}}\ and\ \bibinfo {author} {\bibfnamefont {F.}~\bibnamefont
  {Marquardt}},\ }\href {\doibase 10.1103/PhysRevLett.111.133601} {\bibfield
  {journal} {\bibinfo  {journal} {Phys. Rev. Lett.}\ }\textbf {\bibinfo
  {volume} {111}},\ \bibinfo {pages} {133601} (\bibinfo {year}
  {2013})}\BibitemShut {NoStop}%
\bibitem [{\citenamefont {Rabl}(2011)}]{Rabl2011}%
  \BibitemOpen
  \bibfield  {author} {\bibinfo {author} {\bibfnamefont {P.}~\bibnamefont
  {Rabl}},\ }\href {\doibase 10.1103/PhysRevLett.107.063601} {\bibfield
  {journal} {\bibinfo  {journal} {Phys. Rev. Lett.}\ }\textbf {\bibinfo
  {volume} {107}},\ \bibinfo {pages} {063601} (\bibinfo {year}
  {2011})}\BibitemShut {NoStop}%
\bibitem [{\citenamefont {Nunnenkamp}\ \emph {et~al.}(2011)\citenamefont
  {Nunnenkamp}, \citenamefont {B\o{}rkje},\ and\ \citenamefont
  {Girvin}}]{Nunnenkamp11}%
  \BibitemOpen
  \bibfield  {author} {\bibinfo {author} {\bibfnamefont {A.}~\bibnamefont
  {Nunnenkamp}}, \bibinfo {author} {\bibfnamefont {K.}~\bibnamefont
  {B\o{}rkje}}, \ and\ \bibinfo {author} {\bibfnamefont {S.~M.}\ \bibnamefont
  {Girvin}},\ }\href {\doibase 10.1103/PhysRevLett.107.063602} {\bibfield
  {journal} {\bibinfo  {journal} {Phys. Rev. Lett.}\ }\textbf {\bibinfo
  {volume} {107}},\ \bibinfo {pages} {063602} (\bibinfo {year}
  {2011})}\BibitemShut {NoStop}%
\bibitem [{\citenamefont {Qian}\ \emph {et~al.}(2012)\citenamefont {Qian},
  \citenamefont {Clerk}, \citenamefont {Hammerer},\ and\ \citenamefont
  {Marquardt}}]{Qian12}%
  \BibitemOpen
  \bibfield  {author} {\bibinfo {author} {\bibfnamefont {J.}~\bibnamefont
  {Qian}}, \bibinfo {author} {\bibfnamefont {A.~A.}\ \bibnamefont {Clerk}},
  \bibinfo {author} {\bibfnamefont {K.}~\bibnamefont {Hammerer}}, \ and\
  \bibinfo {author} {\bibfnamefont {F.}~\bibnamefont {Marquardt}},\ }\href
  {\doibase 10.1103/PhysRevLett.109.253601} {\bibfield  {journal} {\bibinfo
  {journal} {Phys. Rev. Lett.}\ }\textbf {\bibinfo {volume} {109}},\ \bibinfo
  {pages} {253601} (\bibinfo {year} {2012})}\BibitemShut {NoStop}%
\bibitem [{\citenamefont {K\'om\'ar}\ \emph {et~al.}(2013)\citenamefont
  {K\'om\'ar}, \citenamefont {Bennett}, \citenamefont {Stannigel},
  \citenamefont {Habraken}, \citenamefont {Rabl}, \citenamefont {Zoller},\ and\
  \citenamefont {Lukin}}]{Komar13}%
  \BibitemOpen
  \bibfield  {author} {\bibinfo {author} {\bibfnamefont {P.}~\bibnamefont
  {K\'om\'ar}}, \bibinfo {author} {\bibfnamefont {S.~D.}\ \bibnamefont
  {Bennett}}, \bibinfo {author} {\bibfnamefont {K.}~\bibnamefont {Stannigel}},
  \bibinfo {author} {\bibfnamefont {S.~J.~M.}\ \bibnamefont {Habraken}},
  \bibinfo {author} {\bibfnamefont {P.}~\bibnamefont {Rabl}}, \bibinfo {author}
  {\bibfnamefont {P.}~\bibnamefont {Zoller}}, \ and\ \bibinfo {author}
  {\bibfnamefont {M.~D.}\ \bibnamefont {Lukin}},\ }\href {\doibase
  10.1103/PhysRevA.87.013839} {\bibfield  {journal} {\bibinfo  {journal} {Phys.
  Rev. A}\ }\textbf {\bibinfo {volume} {87}},\ \bibinfo {pages} {013839}
  (\bibinfo {year} {2013})}\BibitemShut {NoStop}%
\bibitem [{\citenamefont {Lu}\ \emph {et~al.}(2013)\citenamefont {Lu},
  \citenamefont {Zhang}, \citenamefont {Ashhab}, \citenamefont {Wu},\ and\
  \citenamefont {Nori}}]{Lu2013}%
  \BibitemOpen
  \bibfield  {author} {\bibinfo {author} {\bibfnamefont {X.-Y.}\ \bibnamefont
  {Lu}}, \bibinfo {author} {\bibfnamefont {W.-M.}\ \bibnamefont {Zhang}},
  \bibinfo {author} {\bibfnamefont {S.}~\bibnamefont {Ashhab}}, \bibinfo
  {author} {\bibfnamefont {Y.}~\bibnamefont {Wu}}, \ and\ \bibinfo {author}
  {\bibfnamefont {F.}~\bibnamefont {Nori}},\ }\href {\doibase
  10.1038/srep02943} {\bibfield  {journal} {\bibinfo  {journal} {Sci. Rep.}\
  }\textbf {\bibinfo {volume} {3}},\ \bibinfo {pages} {2943} (\bibinfo {year}
  {2013})}\BibitemShut {NoStop}%
\bibitem [{\citenamefont {Xu}\ and\ \citenamefont {Taylor}(2013)}]{Xu2013}%
  \BibitemOpen
  \bibfield  {author} {\bibinfo {author} {\bibfnamefont {X.}~\bibnamefont
  {Xu}}\ and\ \bibinfo {author} {\bibfnamefont {J.~M.}\ \bibnamefont
  {Taylor}},\ }\href {http://arxiv.org/abs/1303.7469} {} (\bibinfo {year}
  {2013}),\ \Eprint {http://arxiv.org/abs/arXiv:1303.7469} {arXiv:1303.7469}
  \BibitemShut {NoStop}%
\bibitem [{sup()}]{supp}%
  \BibitemOpen
  \href@noop {} {}\bibinfo {note} {Supplementary material contains the
  derivation of the full normal mode Hamiltonian and the derivation of the
  minimum $g^{(2)}(0)$ when quantum jumps are neglected.}\BibitemShut {Stop}%
\bibitem [{\citenamefont {Chan}\ \emph {et~al.}(2009)\citenamefont {Chan},
  \citenamefont {Eichenfield}, \citenamefont {Camacho},\ and\ \citenamefont
  {Painter}}]{Chan09}%
  \BibitemOpen
  \bibfield  {author} {\bibinfo {author} {\bibfnamefont {J.}~\bibnamefont
  {Chan}}, \bibinfo {author} {\bibfnamefont {M.}~\bibnamefont {Eichenfield}},
  \bibinfo {author} {\bibfnamefont {R.}~\bibnamefont {Camacho}}, \ and\
  \bibinfo {author} {\bibfnamefont {O.}~\bibnamefont {Painter}},\ }\href
  {\doibase 10.1364/OE.17.003802} {\bibfield  {journal} {\bibinfo  {journal}
  {Opt. Express}\ }\textbf {\bibinfo {volume} {17}},\ \bibinfo {pages} {3802}
  (\bibinfo {year} {2009})}\BibitemShut {NoStop}%
\bibitem [{\citenamefont {Chan}\ \emph {et~al.}(2012)\citenamefont {Chan},
  \citenamefont {Safavi-Naeini}, \citenamefont {Hill}, \citenamefont
  {Meenehan},\ and\ \citenamefont {Painter}}]{Chan12}%
  \BibitemOpen
  \bibfield  {author} {\bibinfo {author} {\bibfnamefont {J.}~\bibnamefont
  {Chan}}, \bibinfo {author} {\bibfnamefont {A.~H.}\ \bibnamefont
  {Safavi-Naeini}}, \bibinfo {author} {\bibfnamefont {J.~T.}\ \bibnamefont
  {Hill}}, \bibinfo {author} {\bibfnamefont {S.}~\bibnamefont {Meenehan}}, \
  and\ \bibinfo {author} {\bibfnamefont {O.}~\bibnamefont {Painter}},\ }\href
  {\doibase http://dx.doi.org/10.1063/1.4747726} {\bibfield  {journal}
  {\bibinfo  {journal} {Applied Physics Letters}\ }\textbf {\bibinfo {volume}
  {101}},\ \bibinfo {eid} {081115} (\bibinfo {year} {2012})}\BibitemShut
  {NoStop}%
\bibitem [{\citenamefont {Gomis-Bresco}\ \emph {et~al.}(2014)\citenamefont
  {Gomis-Bresco}, \citenamefont {Navarro-Urrios}, \citenamefont {Oudich},
  \citenamefont {El-Jallal}, \citenamefont {Griol}, \citenamefont {Puerto},
  \citenamefont {Chavez}, \citenamefont {Pennec}, \citenamefont
  {Djafari-Rouhani}, \citenamefont {Alzina}, \citenamefont {Mart地ez},\ and\
  \citenamefont {Torres}}]{Gomis-Bresco14}%
  \BibitemOpen
  \bibfield  {author} {\bibinfo {author} {\bibfnamefont {J.}~\bibnamefont
  {Gomis-Bresco}}, \bibinfo {author} {\bibfnamefont {D.}~\bibnamefont
  {Navarro-Urrios}}, \bibinfo {author} {\bibfnamefont {M.}~\bibnamefont
  {Oudich}}, \bibinfo {author} {\bibfnamefont {S.}~\bibnamefont {El-Jallal}},
  \bibinfo {author} {\bibfnamefont {A.}~\bibnamefont {Griol}}, \bibinfo
  {author} {\bibfnamefont {D.}~\bibnamefont {Puerto}}, \bibinfo {author}
  {\bibfnamefont {E.}~\bibnamefont {Chavez}}, \bibinfo {author} {\bibfnamefont
  {Y.}~\bibnamefont {Pennec}}, \bibinfo {author} {\bibfnamefont
  {B.}~\bibnamefont {Djafari-Rouhani}}, \bibinfo {author} {\bibfnamefont
  {F.}~\bibnamefont {Alzina}}, \bibinfo {author} {\bibfnamefont
  {A.}~\bibnamefont {Mart地ez}}, \ and\ \bibinfo {author} {\bibfnamefont
  {C.~S.}\ \bibnamefont {Torres}},\ }\href {\doibase 10.1038/ncomms5452}
  {\bibfield  {journal} {\bibinfo  {journal} {Nat Commun}\ }\textbf {\bibinfo
  {volume} {5}},\ \bibinfo {pages} {4452} (\bibinfo {year} {2014})}\BibitemShut
  {NoStop}%
\bibitem [{\citenamefont {Safavi-Naeini}\ \emph {et~al.}(2014)\citenamefont
  {Safavi-Naeini}, \citenamefont {Hill}, \citenamefont {Meenehan},
  \citenamefont {Chan}, \citenamefont {Gr\"oblacher},\ and\ \citenamefont
  {Painter}}]{Safavi-Naeini14}%
  \BibitemOpen
  \bibfield  {author} {\bibinfo {author} {\bibfnamefont {A.~H.}\ \bibnamefont
  {Safavi-Naeini}}, \bibinfo {author} {\bibfnamefont {J.~T.}\ \bibnamefont
  {Hill}}, \bibinfo {author} {\bibfnamefont {S.}~\bibnamefont {Meenehan}},
  \bibinfo {author} {\bibfnamefont {J.}~\bibnamefont {Chan}}, \bibinfo {author}
  {\bibfnamefont {S.}~\bibnamefont {Gr\"oblacher}}, \ and\ \bibinfo {author}
  {\bibfnamefont {O.}~\bibnamefont {Painter}},\ }\href {\doibase
  10.1103/PhysRevLett.112.153603} {\bibfield  {journal} {\bibinfo  {journal}
  {Phys. Rev. Lett.}\ }\textbf {\bibinfo {volume} {112}},\ \bibinfo {pages}
  {153603} (\bibinfo {year} {2014})}\BibitemShut {NoStop}%
\bibitem [{\citenamefont {Sekoguchi}\ \emph {et~al.}(2014)\citenamefont
  {Sekoguchi}, \citenamefont {Takahashi}, \citenamefont {Asano},\ and\
  \citenamefont {Noda}}]{Sekoguchi14}%
  \BibitemOpen
  \bibfield  {author} {\bibinfo {author} {\bibfnamefont {H.}~\bibnamefont
  {Sekoguchi}}, \bibinfo {author} {\bibfnamefont {Y.}~\bibnamefont
  {Takahashi}}, \bibinfo {author} {\bibfnamefont {T.}~\bibnamefont {Asano}}, \
  and\ \bibinfo {author} {\bibfnamefont {S.}~\bibnamefont {Noda}},\ }\href
  {\doibase 10.1364/OE.22.000916} {\bibfield  {journal} {\bibinfo  {journal}
  {Opt. Express}\ }\textbf {\bibinfo {volume} {22}},\ \bibinfo {pages} {916}
  (\bibinfo {year} {2014})}\BibitemShut {NoStop}%
\end{thebibliography}%

\begin{widetext}
\clearpage
\begin{center}
{\Large Supplemental Material: \\ \vspace{8pt}
 Quantum nonlinear optics near optomechanical instabilities}
 \end{center}
 \vspace{20pt}
\end{widetext}

\renewcommand{\thesection}{S\arabic{section}}   
\renewcommand{\thetable}{S\arabic{table}}   
\renewcommand{\thefigure}{S\arabic{figure}}
\renewcommand{\theequation}{S\arabic{equation}}
\setcounter{page}{1}
\setcounter{equation}{0}

\section{Diagonalization of the bilinear hamiltonian}
The bilinear hamiltonian is
\begin{align}
H_0 = \Delta_a a^{\dagger}a + \Delta_b b^{\dagger}b + \omega_m c^{\dagger}c -  G_0 (b+b^{\dagger}) (c^{\dagger} + c).
\end{align}
In this hamiltonian, mode $a$ is already decoupled, so we only need to diagonalize the coupled harmonic oscillator subsystem $b-c$. Define 
\begin{subequations}
\begin{align}
X_b & = (b+b^{\dagger})/\sqrt{2}, \quad Y_b = (b-b^{\dagger})/i\sqrt{2}, \\
X_c & = (c+c^{\dagger})/\sqrt{2}, \quad  Y_c = (c-c^{\dagger})/i\sqrt{2}.
\end{align}
\end{subequations}
They satisfy the commutation relations $[X_b, Y_b] = [X_c, Y_c] = i$, $[X_b, X_c] = [X_b, Y_c] = [Y_b, X_c] = [Y_b, Y_c] = 0$. We can then rewrite the hamiltonian of the $b-c$ subsystem as  
\begin{equation}
H_{bc}= \frac{1}{2}\Delta_b (X_b^2 + Y_b^2) + \frac{1}{2}\omega_m (X_c^2 + Y_c^2) - 2 G_0  X_b X_c. 
\end{equation}
We now rescale the operators  $X_c$ and $Y_c$ according to  
\begin{equation}
X_c = X_c^{\prime} \sqrt{\omega_m/\Delta_b},  \quad Y_c =  Y_c^{\prime} \sqrt{\Delta_b/\omega_m},
\end{equation}
but keep $X_b$ and $Y_b$ the same 
\begin{equation}
X_b = X_b^{\prime} ,  \quad Y_b =  Y_b^{\prime}. 
\end{equation}
In this transformed basis the hamiltonian is 
\begin{align}
H_{bc} & = \frac{1}{2}\Delta_b ({X_b^{\prime}}^2 + {Y_b^{\prime}}^2) + \frac{1}{2}\Delta_b \left(\frac{\omega_m^2}{\Delta_b^2}{X_c^{\prime}}^2 + {Y_c^{\prime}}^2\right) \nonumber \\
& - 2 G_0 \sqrt{\frac{\omega_m}{\Delta_b}} X_b^{\prime} X_c^{\prime}. 
\end{align}
We then make a unitary transformation to get the normal mode coordinates that yields 
\begin{subequations}
\begin{align}
& \left( \begin{array}{c}
X_b^{\prime} \\
X_c^{\prime} \\
\end{array}   \right) 
= \left( \begin{array}{cc}
\alpha & \beta \\
-\beta & \alpha \\
\end{array} \right)
\left( \begin{array}{c}
X_+ \\
X_- \\
\end{array} \right),  \\
&\left( \begin{array}{c}
Y_b^{\prime} \\
Y_c^{\prime} \\
\end{array}   \right) 
= \left( \begin{array}{cc}
\alpha & \beta \\
-\beta & \alpha \\
\end{array} \right)
\left( \begin{array}{c}
Y_+ \\
Y_- \\
\end{array}   \right). 
\end{align}
\end{subequations}
The commutation relations are preserved if $\alpha^2 + \beta^2 = 1$ ($\alpha, \beta$ are real). 
\begin{widetext}
So the hamiltonian of the $b-c$ subsystem is given by 
\begin{align}
H_{bc}  =& \frac{1}{2}\Delta_b \left ( \alpha^2  + \frac{\omega_m^2}{\Delta_b^2} \beta^2 +  \frac{4G_0}{\Delta_b}\sqrt{\frac{\omega_m}{\Delta_b}} \alpha\beta \right)  X_+^2 + \frac{1}{2}\Delta_b \left ( \beta^2  + \frac{\omega_m^2}{\Delta_b^2} \alpha^2 -  \frac{4G_0}{\Delta_b} \sqrt{\frac{\omega_m}{\Delta_b}} \alpha\beta \right)  X_-^2 \nonumber \\
&+  \frac{1}{2}\Delta_b \left(Y_+^2 + Y_-^2\right) +  \left[\frac{1}{2}\Delta_b 2\alpha\beta (1-\frac{\omega_m^2}{\Delta_b^2}) -  2 G_0 \sqrt{\frac{\omega_m}{\Delta_b}} (\alpha^2 - \beta^2) \right ]X_+ X_-. 
\end{align}
\end{widetext}
It is diagonal if the cross term $X_+ X_-$ is zero,
\begin{equation}
\Delta_b \alpha\beta (1-\frac{\omega_m^2}{\Delta_b^2}) - 2G_0 \sqrt{\frac{\omega_m}{\Delta_b}} (\alpha^2 - \beta^2) = 0. 
\end{equation}
This condition along with  $\alpha^2 + \beta^2 =1$  determines $\alpha$ and $\beta$ for the normal modes. The diagonalized hamiltonian thus becomes 
\begin{align}
H_{bc}  =& \frac{1}{2}\Delta_b( \xi_+^2 X_+^2 + Y_+^2) +  \frac{1}{2}\Delta_b (\xi_-^2 X_-^2 + Y_-^2) 
\end{align}
with 
\begin{subequations}
\begin{align}
\xi_+^2 &= \alpha^2  + \frac{\omega_m^2}{\Delta_b^2} \beta^2 +  \frac{4G_0}{\Delta_b}\sqrt{\frac{\omega_m}{\Delta_b}} \alpha\beta , \\
\xi_-^2 &= \beta^2  + \frac{\omega_m^2}{\Delta_b^2} \alpha^2 -  \frac{4G_0}{\Delta_b}\sqrt{\frac{\omega_m}{\Delta_b}} \alpha\beta .
\end{align}
\end{subequations}
In the limit described in the main text with $\omega_m \ll \Delta_b$, $\xi_+$ is approximately one and $\xi_- \approx  \eta \zeta$.

This hamiltonian describes two decoupled harmonic oscillators $\mathrm{HO}+$ and $\mathrm{HO}-$ with effective masses $m_{\pm} = \Delta_b^{-1}$ and effective frequencies $\omega_{\pm} = \Delta_b\xi_{\pm}$, so the hamiltonian can be rewritten as 
\begin{align}
H_{bc} & = H_{+} + H_{-}  \\
& = \frac{Y_+^2}{2\Delta_b^{-1}} + \frac{1}{2}\Delta_b^{-1} (\Delta_b \xi_+)^2 X_+^2 \nonumber \\
&  + \frac{Y_-^2}{2\Delta_b^{-1}} + \frac{1}{2}\Delta_b^{-1} (\Delta_b \xi_-)^2 X_-^2 .  \nonumber 
\end{align}
 We can write the wavefunction of the $n{\mathrm{th}}$ eigenstate of $\mathrm{HO}-$ (for example)  in position representation:
\begin{equation}
\psi_n(X_-) = \frac{1}{\sqrt{2^n n!}} \left(\frac{\xi_-}{\pi}\right)^{1/4} e^{-\frac{\xi_-}{2} X_-^2} H_n(\sqrt{\xi_-} X_- ). 
\end{equation}

\section{Hamiltonian in the normal mode basis}
We now define new squeezed operators
\begin{equation}
d_{\pm} = \sqrt{\frac{\xi_{\pm}}{2}} X_{\pm} + i \sqrt{\frac{1}{2\xi_{\pm}}} Y_{\pm}
\end{equation}
so $[d_{\pm}, d_{\pm}^{\dagger}] = 1$, $[d_\pm,d_{\mp}^\dagger]=0$, and $\xi_{\pm}^2 X_{\pm}^2 + Y_{\pm}^2 = 2\xi_{\pm} (d_{\pm}^{\dagger}d_{\pm} + 1)$.  The bilinear hamiltonian written in new operators is 
\begin{equation}
H_0  = \Delta_a a^{\dagger}a +  \omega_+ d_+^{\dagger}d_+ +    \omega_- d_-^{\dagger}d_-. 
\end{equation}
The normal mode operators written in original operators are: 
\begin{widetext}
\begin{subequations} 
\begin{align}
d_+ & = \frac{\sqrt{\xi_+}}{2} \left[\alpha (b+b^{\dagger}) -\beta \sqrt{\frac{\Delta_b}{\omega_m}} (c+c^{\dagger}) \right] +  \frac{1}{2\sqrt{\xi_+}}  \left[\alpha (b-b^{\dagger})  - \beta\sqrt{\frac{\omega_m}{\Delta_b}} (c - c^{\dagger})\right], \\
d_- & = \frac{\sqrt{\xi_-}}{2} \left[\beta  (b+b^{\dagger}) + \alpha\sqrt{\frac{\Delta_b}{\omega_m}} (c+c^{\dagger}) \right] +  \frac{1}{2\sqrt{\xi_-}}  \left[ \beta  (b-b^{\dagger})+ \alpha \sqrt{\frac{\omega_m}{\Delta_b}} (c - c^{\dagger})\right]. 
\end{align}
\end{subequations}
and the inverse:
\begin{subequations}
\begin{align}
b & = \frac{\alpha}{2} \left[ \frac{1}{\sqrt{\xi_+}} (d_+ +d_+^{\dagger} )  + \sqrt{\xi_+} (d_+ - d_+^{\dagger} ) \right] +   \frac{\beta}{2} \left[ \frac{1}{\sqrt{\xi_-}}  (d_- +d_-^{\dagger})  + \sqrt{\xi_-}  (d_- - d_-^{\dagger}) \right], \\
c & = - \frac{\beta}{2} \left[  \sqrt{\frac{\omega_m}{\xi_+\Delta_b}} (d_+ + d_+^{\dagger})   +   \sqrt{\frac{\xi_+\Delta_b}{\omega_m}} (d_+ - d_+^{\dagger})   \right] +  \frac{\alpha}{2} \left[ \sqrt{\frac{\omega_m}{\xi_-\Delta_b}}(d_- + d_-^{\dagger}) +  \sqrt{\frac{\xi_-\Delta_b}{\omega_m}} (d_- - d_-^{\dagger})   \right]  .
\end{align}
\end{subequations}
Now it is straightforward to write the nonlinear coupling in terms of the normal mode coordinates:
\begin{align}
H_{\mathrm{nl}} & = - g_0 a^{\dagger} b(c+c^{\dagger}) + \mathrm{h.c.}  \nonumber  \\
& = -  g_0 \sqrt{\frac{\omega_m}{\Delta_b}} 
\left[ -\frac{\alpha\beta}{2\xi_+} (a+a^{\dagger}) (d_+ + d_+^{\dagger})^2 + \frac{\alpha\beta}{2}(a-a^{\dagger}) (d_+^2  - d_+^{\dagger 2}) \right.  \\
& \quad \left. + \frac{\alpha^2 -\beta^2}{2\sqrt{\xi_+ \xi_-}} (a+a^{\dagger}) (d_+ + d_+^{\dagger}) (d_- + d_-^{\dagger}) + \frac{\beta^2}{2}\sqrt{\frac{\xi_-}{\xi_+}} (a-a^{\dagger}) (d_+ + d_+^{\dagger})(d_- - d_-^{\dagger}) \right.  \nonumber \\
& \quad \left. - \frac{\alpha^2}{2} \sqrt{\frac{\xi_+}{\xi_-}} (a-a^{\dagger}) (d_- + d_-^{\dagger})(d_+ - d_+^{\dagger}) + \frac{\alpha\beta}{2\xi_-} (a+a^{\dagger}) (d_- + d_-^{\dagger})^2 - \frac{\alpha\beta}{2} (a-a^{\dagger}) (d_-^2 - d_-^{\dagger 2})  \right] .  \nonumber 
\end{align}
\end{widetext}

Define $ \tan\phi = \cfrac{\omega_m}{\Delta_b}, ~ r = 2\cfrac{G_0}{\omega_m} \sqrt{\cfrac{\omega_m}{\Delta_b}}$ and  $\alpha = \cos\theta, ~\beta = \sin\theta$, then the diagonalization condition reduces to 
\begin{equation}
\tan2\theta = r \tan 2\phi, 
\end{equation}
and the normal mode energies become
\begin{equation}
\xi_{\pm}^2 =  \frac{1}{2} ( 1 + \tan^2\phi) \left( 1\pm \sqrt{\cos^2 2\phi + r^2 \sin^2 2\phi} \right). 
\end{equation} 
We now consider the regime where the mechanical frequency is small compared to the detuning of mode $b$ and the driving is so strong that $r$ is close to 1. This allows us to introduce two small parameters $\eta \equiv \omega_m/\Delta_b$ and $\zeta \equiv \sqrt{1-r^2}$. When $\eta \ll 1$, $\tan\phi \approx \sin\phi  \ll1$, and we have
\begin{subequations}
\begin{align}
\xi_+ &\approx 1 + r^2\omega_m^2/2\Delta_b^2 = 1 + r^2\eta^2/2, \\
\xi_-  &\approx  \sqrt{1-r^2}\omega_m/\Delta_b = \zeta\eta. 
\end{align}
\end{subequations}
The diagonalized hamiltonian becomes:
\be
H_0 = \Delta_a a^{\dag}a + (\Delta_b + \delta)\bar{b}^{\dag} \bar{b} + \omega_m\zeta d^{\dag} d, 
\ee
with $\delta = r^2\omega_m\eta/2$ and the new notations for the normal modes are defined as: 
\begin{subequations}
\begin{align}
\bar{b} & \equiv d_+ \approx  b - \frac{r}{2}\sqrt{\eta} (c+c^\dagger), \\ 
d & \equiv d_- \approx \frac{1}{2 \sqrt{\zeta}}(c-c^\dagger)+ \frac{\sqrt{\zeta}}{2} (c+c^\dagger)   + \frac{r}{2}\sqrt{\frac{\eta}{\zeta}}(b-b^\dagger). 
\end{align}
\end{subequations}

\section{Derivation of $g^2(0)$ when quantum jumps are neglected}
Here we show the standard procedure for calculating the two-photon correlation function $g^2(0)$ in the quasi-steady state regime using an effective hamiltonian. We consider the hamiltonian Eq.~(1) (main text) with antihermitian terms describing the dissipation and weak coherent probe field on the $\bar{b}$ mode at frequency $\omega_p = \Delta_{\bar{b}} - g_{\mathrm{nl}}$: 
\begin{align} 
H_{\textrm{eff}} & = (\Delta_a - i\kappa/2)  a^{\dagger}a + (\Delta_{\bar{b}} - i\kappa/2) \bar{b}^{\dagger}\bar{b} + \omega_m\zeta d^{\dagger}d \nonumber \\
& - g_{\textrm{nl}}(a^{\dagger}\bar{b} +  \bar{b}^{\dagger} a)(d+d^{\dagger}) + i\beta_{\bar{b}} (\bar{b}^{\dagger} e^{-i\omega_p t} - \bar{b}e^{i\omega_pt} ). 
\end{align}
The term $(a+a^{\dagger})(d^2 + d^{\dagger 2} + 2d^{\dagger}d)$ has been neglected since its strength is weak in the limit $\Delta_b \gg \omega_m$ and it is also far off resonant. Moving to a frame rotating at $\omega_p $ for the optical fields and using the resonance condition $\Delta_{\bar{b}} = \Delta_a + \omega_m\zeta$, we get 
\begin{align} 
H_{\textrm{eff}} &= (-\omega_m\zeta - i\kappa/2)  a^{\dagger}a + (g_{\mathrm{nl}} - i\kappa/2) \bar{b}^{\dagger}\bar{b} + \omega_m\zeta d^{\dagger}d \nonumber \\
&  - g_{\textrm{nl}}(a^{\dagger}\bar{b} +  \bar{b}^{\dagger} a)(d+d^{\dagger}) + i\beta_{\bar{b}} (\bar{b}^{\dagger} - \bar{b}). 
\end{align}

The system evolves according to the effective hamiltonian and we can expand its quasi-steady state in the following basis: 
\begin{align}
\ket{\psi}_{\mathrm{ss}} & = \ket{0,0,0} + c_1\ket{0,1,0}+ c_2\ket{1,0,1}+ c_3\ket{0,1,2} \nonumber \\
& + c_4\ket{0,2,0}+ c_5\ket{1,1,1}+ c_6\ket{2,0,2}+ c_7\ket{0,2,2} \nonumber \\
& + c_8\ket{200}. 
\end{align}
Considering the following coupling between basis states
\begin{align}  
&  \ket{0, 1, 0}\xlongleftrightarrow{ g_{\mathrm{nl}}} \ket{1,0,1} \xlongleftrightarrow{ \sqrt{2}g_{\mathrm{nl}}} \ket{0,1,2}  \nonumber \\
&  \ket{0, 2, 0}\xlongleftrightarrow{\sqrt{2} g_{\mathrm{nl}}} \ket{1,1,1}  \xlongleftrightarrow{2 g_{\mathrm{nl}}} \ket{2,0,2}  \nonumber \\
&\hspace{22.5ex} \xlongleftrightarrow{ 2 g_{\mathrm{nl}}} \ket{0,2,2} \nonumber \\
&\hspace{22.5ex} \xlongleftrightarrow{ \sqrt{2} g_{\mathrm{nl}}} \ket{2,0,0} \nonumber 
\end{align} 
and the pumping processes
\begin{align}
& \ket{0,0,0} \xlongleftrightarrow{\pm i\beta_{\bar{b}}} \ket{0,1,0}  \xlongleftrightarrow{\pm i\beta} \ket{0,2,0}  \nonumber\\
& \ket{1,0,1}  \xlongleftrightarrow{\pm i\beta_{\bar{b}}} \ket{1,1,1}  \nonumber\\
& \ket{0,1,2}  \xlongleftrightarrow{\pm i\beta_{\bar{b}}} \ket{0,2,2}, \nonumber
\end{align}
we can then construct the matrix representation of the effective hamiltonian. 

The steady state is found using the Schrodinger equation:
\begin{equation}
0 = i \frac{\partial}{\partial t} \ket{\psi}_{\mathrm{ss}} = H_{\mathrm{eff}} \ket{\psi}_{\mathrm{ss}}. 
\end{equation}
Solving this set of algebra equations gives us the steady state $\psi_{\mathrm{ss}}$. The $g^2(0)$ is calculated using Eq.~(13) (main text) in the limit $\beta_{\bar{b}} \to 0$.

\end{document}